\documentclass[12pt]{article}
\pdfoutput = 1
\usepackage{graphicx}
\usepackage{adjustbox}
\usepackage{amsmath}
\usepackage{bbm}
\usepackage{fancybox}
\usepackage{amssymb}
\usepackage{amsthm}
\usepackage{amsfonts,amsmath}
\usepackage[section]{placeins}
\usepackage{lineno}
\usepackage{caption}
\usepackage{subcaption}
\usepackage{authblk}
\usepackage{float}
\usepackage{nameref,hyperref}
\usepackage{setspace}  
\usepackage{url}  

\usepackage{verbatim}
\usepackage{indentfirst}
\usepackage{setspace}

\usepackage{xcolor}

\newtheorem{theorem}{Theorem}
\newtheorem{proposition}{Proposition}[theorem]

\providecommand{\keywords}[1]
{
  \small	
  \textbf{\textit{Keywords:}} #1
}

\title{Affective Polarization on Small-World and Scale-Free Networks\\
}

\date{}
\author[1]{Alisson Serracín Morales}
\author[2]{Buddhika Nettasinghe}

\affil[1]{Dept. of  Mathematics, Applied Mathematical and Computational Sciences, University of Iowa, Iowa City, IA 52242, USA}
\affil[2]{Dept. of Business Analytics, Tippie College of Business, University of Iowa, Iowa City, IA 52242, USA}

\begin{document}

\maketitle

\begin{abstract}

Affective polarization, the emotional divide characterized by in-group love (trust towards fellow partisans) and out-group hate (mistrust towards those with opposite political views), has become prevalent in the current society. Despite its prevalence, the role of social network structure in the dynamics of affective polarization is yet to be well-understood. We provide a mean-field approximation of opinion dynamics under affective polarization on Watts-Strogatz and power-law (scale-free) networks. 
Our results show that consensus is fragile in social networks with power-law degree distributions, and the smaller average path length of the network (resembling a small-world network) makes achieving the consensus further difficult. Simulations and numerical experiments on real-world networks indicate that the mean-field model is aligned with the actual dynamics. Our findings shed light on how real-world network properties shape the dynamics of affective polarization and why consensus remains elusive in the real-world.

\end{abstract}

\keywords{social networks, affective polarization, consensus, partisan polarization, opinion dynamics, power-law, Watts-Strogatz model}

\maketitle

\singlespacing
\section{Introduction} \label{sec:intro}
\onehalfspacing

Affective polarization, the emotional divide characterized by in-group love (trust towards fellow partisans) and out-group hate (mistrust towards those with opposite political views), has  become more prevalent in the world \cite{phillips2022affective}. It has grown particularly noticeable in American society, \cite{10.1093/pnasnexus/pgae310} where Democrats and Republicans tend to view people in the opposite party negatively (e.g.,~hypocritical, selfish, closed-minded) and are opposed to interacting positively across party lines \cite{APintro}, \cite{Affectnotideol}, \cite{Whitt2021}. Consequently, this phenomenon affects how people form opinions and make decisions based on those of others that they observe in their own party as well as the opposing party \cite{RudolphAP}. 
For example, people's choices such as masking were polarized along ideological lines in the US during the COVID-19 Pandemic \cite{druckman2020affective}, \cite{GreenCOVIDAP}, \cite{GuyCOVID}. Similar dynamics have also been observed when high-profile executives of companies publicly support or oppose political candidates, leading to ideologically polarized shifts in the public perception of the company's products  and services~\cite{GAO20251248, green2025tesla}, \cite{sumner2022cost}. 

In this work, our aim is to understand the implications of social network structure on such dynamics. More specifically, we explore how different network structures can shape the dynamics of people's choices in an affectively polarized society with a particular focus on scale-free (power-law) networks. Our work builds up on \cite{nettasinghe2025outgroup}  which proposed a mathematical model of how people make binary choices (e.g., wear a mask or not, get vaccinated or not, drink matcha or coffee) in an affectively polarized social network containing two political parties (red and blue). In the model proposed in \cite{nettasinghe2025outgroup} (and  generalized subsequently in \cite{nettasinghe2025group}), people's decisions are shaped by the degree to which they like and trust the members of their own group (in-group love) and dislike and distrust towards members of the other group (out-group hate). \cite{nettasinghe2025outgroup} analyzes the model dynamics in fully-connected networks and stochastic block models showing that either consensus (all individuals adopting the same choice), polarization along party-lines (each group adopting opposite choices) or non-partisan polarization (where each group decisions gets uniformly split between two choices)  eventually emerge. The final outcome is determined by three key factors: political group composition in the society (red-blue ratio), in-group love to out-group hate ratio, and the level of homophily. These findings show how the dynamics of choices are shaped by both real-world network properties (e.g.,~homophily, group sizes) as well as characteristic features of affective polarization (in-group love and out-group hate). 

Although the model presented in \cite{nettasinghe2025outgroup} can be implemented on any graph, so far, its dynamics have been theoretically analyzed only for fully connected graphs and stochastic block models. The implications of other well-documented properties of real-world networks (that are not captured by those two models) on the dynamics of affective polarization have not been well-studied yet. For example, \emph{How hard is it to achieve consensus in real-world networks that often have power-law (scale-free) degree distributions? How does the shorter average path length (small-world property) shape its dynamics?} 
To fill this gap, we extend the analysis of the model proposed in \cite{nettasinghe2025outgroup} to understand how different network structures lead to different outcomes (e.g., consensus, polarization, etc.) in an affectively polarized society. In particular, we make the following contributions:

\begin{itemize}
    \item We derive closed-form mean-field dynamics of opinion dynamics for Watts-Strogatz model~\cite{watts1998small} and networks with power-law degree distributions~(via configuration model)~\cite{babarasi1999}. The closed-form reduces the need for computationally intensive Monte-Carlo simulations and facilitates theoretical analysis.

    \item We provide numerical experiments on synthetic networks as well as real-world networks to illustrate the closer alignment of mean-field dynamics and the actual dynamics. Further, we find that power-law degree distributions lead to much richer variety of emergent behaviors that resemble real-world patterns.
    
    \item Theoretical insights are derived to show the conditions under which perfect consensus, perfect polarization, and states in between emerge depending on the model parameters. In particular, we find that heavy tails in power-law degree distributions and shorter average path lengths can make the consensus difficult to achieve in social networks.    

\end{itemize}

\vspace{0.1cm}
\noindent
{\bf Organization: }The rest of this paper is organized as follows:  Sec.~\ref{sec:related_work} presents related work and preliminaries.  Sec.~\ref{sec:SWModel} and Sec.~\ref{sec:ConfModel} present the mean-field closed-form of the opinion dynamics under affective polarization for Watts-Strogatz networks and configuration model (with power-law degree distributions), respectively. Sec.~\ref{sec:experiments} presents experiments of the model with real world datasets. Finally, Sec.~\ref{sec:discussion} provides a discussion of the results and concluding remarks.

\section{Related Work and Preliminaries}
\label{sec:related_work}

In this section, we discuss related literature at the intersection of network science and affective polarization as well as the preliminaries needed for our subsequent results.

\subsection{Affective Polarization on Networks}
  Some recent work in the literature has taken an explicitly network-centric view of affective polarization.

\cite{UnpackingPolarization} shows a computational method called FAULTANA, that measures the degree of antagonism and alignment in online debates, allowing to study affective polarization in online social networks. The authors applied FAULTANA to Birdwatch, a Twitter fact-checking community and the discussion forums of DerStandard, an Austrian online newspaper. For both online platforms, they found that communities are divided into two large groups that are built on political identities. Along a similar line, \cite{Heatedconversationsinawarmingworld} develops a methodological framework that uses machine learning and network science tools to measure the degree to which stance groups express more negativity towards the out-group than the in-group. Applying the framework to Twitter discourses related to climate change issues, the authors found that disbelievers of climate change displayed more hostility towards the out-group, while believers exhibited less affective polarization. 

In a related vein, \cite{Hohmann2025} introduces a network-based metric of affective polarization that captures both hostility and social distance. The metric was tested on a Twitter data set on COVID-19, where affective polarization was found to remain low in February 2020 and increased in the coming months. \cite{Lerman2024} also focuses on measuring affective polarization on social media. The authors measured affective polarization by analyzing the emotions and toxicity of reply interactions. Results showed that interactions between users with the same ideology are generally positive and interactions between users with an opposite ideology exhibit negativity and toxicity. Moreover, emotions vary with network distance between users. Interactions between users closer to each other tend to reflect positive emotions, while interactions between more distant users lead to  anger, disgust and toxicity.

Furthermore, \cite{MeasuringAPIEEE} estimates affective polarization in social networks using sentiment analysis techniques. The framework was evaluated using data from the regional elections in Spain, specifically in the Region of Murcia. Their findings showed that every political group is polarized in different degrees. Finally, \cite{polbackfire} introduces an opinion dynamics model that suggests that promoting nonlocal interactions often leads to partisan conflicts. This agrees with our finding that shorter average path lengths facilitate party-line polarization.

Thus, the key focus of the literature has been on machine learning and data-driven methods to quantify affective polarization. In contrast, our aim is to understand the interplay between network structure and dynamics of opinions in affectively polarized society. In addition, our results also provide tools for network and computational social scientists to explore affective polarization without the need of intensive simulations.

\subsection{Dynamics of Choices under Affective Polarization}
   Our work builds up on the model of affective polarization presented in \cite{nettasinghe2025outgroup}, which we briefly recap here along with a summary of the main findings. In the model presented in \cite{nettasinghe2025outgroup}, people are connected by a social network and each individual belongs to one of the two groups (e.g.,~conservative and liberal). People like and trust members of their own group (in-group love) and dislike and distrust members of the other group (out-group hate). When faced with two possible choices (e.g.,~mask or not mask), an individual examines choices of their social-connections and attempts to align with the choices of their in-group and deviate from the choices made by members of their out-group.

\vspace{0.1cm}
\noindent
{\bf Model: } Formally, let  $G = (V,E)$ be an undirected social network with $|V| = N$ individuals. There are two binary attributes for each individual $v \in V$: a time-varying binary attribute $H_k(v) \in \{0, 1\}$ that represents $v's$ choice at time k and a time-invariant binary attribute $R(v) \in \{0, 1\}$ representing the group (i.e. political affiliation). We say that node $v$ is red ($v \in \mathcal{R}$) if $R(v)=1$, and node $v$ is blue ($v \in \mathcal{B}$) if $R(v)=0$. The values $N^\mathcal{B} = |\mathcal{B}|$ and $N^\mathcal{R} = |\mathcal{R}|$ denote the number of blue and red nodes, respectively, and $r = \frac{N^\mathcal{R}}{N}$ denotes the fraction of red nodes in the social network. At each time step $k$ (where $k=0,1,2, ...$), a node $X_k \in V$ is selected uniformly at random and updates its state based on the choices of its neighbors. Let
\begin{equation}
d_k^{\text{in}, 0}(X_k) = \sum_{(X_k, u) \in E} \frac{\mathbbm{1} \big(R(u) = R(X_k) \land H_k(u) = 0\big)}{d(X_k)}
    \label{eq:degrees}
\end{equation}

\[
d_k^{\text{in}, 1}(X_k) = \sum_{(X_k, u) \in E} \frac{\mathbbm{1}\big(R(u) = R(X_k) \land H_k(u) = 1\big)}{d(X_k)}
\]

\[
d_k^{\text{out}, 0}(X_k) = \sum_{(X_k, u) \in E} \frac{\mathbbm{1}\big(R(u) \neq R(X_k) \land H_k(u) = 0\big)}{d(X_k)}
\]

\[
d_k^{\text{out}, 1}(X_k) = \sum_{(X_k, u) \in E} \frac{\mathbbm{1}\big(R(u) \neq R(X_k) \land H_k(u) = 1\big)}{d(X_k)},
 \]

represent the number of in-group and out-group neighbors of $X_k$ with choice-0 and choice-1 at time $k$, divided by the total number of neighbors $d(X_k)$. The node $X_k$ updates its choice attribute at time $k+1$ via the rule
\begin{equation}
H_{k+1}(X_k) =
\begin{cases} 
    0 & \text{if } \alpha \big(d_k^{\text{in}, 1}(X_k) - d_k^{\text{in}, 0}(X_k)\big) - \beta \big(d_k^{\text{out}, 1}(X_k) - d_k^{\text{out}, 0}(X_k)\big) < -\delta, \\[1em]
    1 & \text{if } \alpha \big(d_k^{\text{in}, 1}(X_k) - d_k^{\text{in}, 0}(X_k)\big) - \beta \big(d_k^{\text{out}, 1}(X_k) - d_k^{\text{out}, 0}(X_k)\big) > \delta, \\[1em]
    H_k(X_k) & \text{otherwise},
\end{cases}
\label{eq:decision_rule}
\end{equation}
 where $\alpha, \beta, \delta \in [0, 1]$ are constant model parameters.
 The choices of all nodes other than $X_k \in V$ remain unchanged, i.e., for all $u \neq X_k$, $H_{k+1}(u) = H_k(u)$. The parameters $\alpha$ and $\beta$ quantify in-group love and out-group hate, respectively. For example, when $\alpha\ll \beta$, people's decisions depend heavily on what their out-group connections do~i.e.,~decisions are driven more by what the enemies do. On the other hand, $\delta$ quantifies the level of inertia or the minimum social pressure that a person requires to change their choice. A larger $\delta$ value (i.e.,~$\delta\gg0$) implies that a stronger push from the social connections is needed to change the existing choice.

Under the above dynamics, \cite{nettasinghe2025outgroup} analyzes the fraction of nodes in each group (red, blue) that have adopted choice-1 at time $k$~i.e.,~the system's state at time $k$ is given by the column vector ${\theta}_k = [\theta_k^\mathcal{B}, \theta_k^\mathcal{R}]^\prime $ where,
\begin{equation}
\theta_k^\mathcal{B} = \frac{\sum_{v \in V} \mathbbm{1}(R(v) = 0 \land H_k(v) = 1)}{\sum_{v \in V} \mathbbm{1}(R(v) = 0)}, \;\; \theta_k^\mathcal{R} = \frac{\sum_{v \in V}\mathbbm{1}(R(v) = 1 \land H_k(v) = 1)}{\sum_{v \in V} \mathbbm{1}(R(v) = 1)}. 
\label{eqn:discretetheta}
\end{equation}

As $X_k$ is chosen at random, the trajectory of ${\theta}_k = [\theta_k^\mathcal{B}, \theta_k^\mathcal{R}]^\prime$ is also a random process. \cite{nettasinghe2025outgroup} shows that the discrete-time stochastic trajectory of the state of the system $\theta_k, k=0,1,2, ...$  can be approximated using the continuous-time deterministic trajectory of a differential equation (mean-field approximation),
when the graph $G = (V,E)$ is either fully connected or sampled from a stochastic block model. \\
\vspace{0.1cm}
\noindent
{\bf Dynamics under a fully connected network: }
For a fully connected social network $G = (V,E)$, \cite{nettasinghe2025outgroup} obtained the following mean-field approximation for the state of the system for large $N$:

\begin{equation}
\begin{bmatrix}
\dot{\theta}^\mathcal{B} \\
\dot{\theta}^\mathcal{R}
\end{bmatrix}
=
\begin{bmatrix}
(1 - \theta^\mathcal{B})p_\theta^\mathcal{B}(0 \to 1) - \theta^\mathcal{B} p_\theta^\mathcal{B}(1 \to 0) \\
(1 - \theta^\mathcal{R})p_\theta^\mathcal{R}(0 \to 1) - \theta^\mathcal{R} p_\theta^\mathcal{R}(1 \to 0)
\end{bmatrix}, 
\label{eq:systemFC}
\end{equation}
where,
\begin{equation*}
\begin{aligned}
p_{\theta}^\mathcal{B}(0 \to 1) &= \mathbbm{1} \left( \alpha(1 - r)(2\theta^\mathcal{B} - 1) - \beta r(2\theta^\mathcal{R} - 1) > \delta \right), \\
p_{\theta}^\mathcal{B}(1 \to 0) &= \mathbbm{1} \left( \alpha(1 - r)(2\theta^\mathcal{B} - 1) - \beta r(2\theta^\mathcal{R} - 1) < -\delta \right), \\
p_{\theta}^\mathcal{R}(0 \to 1) &= \mathbbm{1} \left( \alpha r(2\theta^\mathcal{R} - 1) - \beta(1 - r)(2\theta^\mathcal{B} - 1) > \delta \right), \\
p_{\theta}^\mathcal{R}(1 \to 0) &= \mathbbm{1} \left( \alpha r(2\theta^\mathcal{R} - 1) - \beta(1 - r)(2\theta^\mathcal{B} - 1) < -\delta \right).
\end{aligned}    
\end{equation*}

The idea behind the Eq.~\ref{eq:systemFC} is that, for a fully connected network, the quantities defined in Eq.~\ref{eq:degrees}{} can be expressed only in terms of the state of the system ${\theta}_k = [\theta_k^\mathcal{B}, \theta_k^\mathcal{R}]^\prime $ and the fraction of the red nodes~$r$.
For example, if $X_k$ picked at random is blue, then,
$$ \mathbb{E}(d_k^{out,1}(X_k))  = r\theta_k^{\mathcal{R}}, \quad \mathbb{E}(d_k^{out,0}(X_k)) = r(1-\theta_k^{\mathcal{R}})$$
$$ \mathbb{E}(d_k^{in,1}(X_k)) = (1-r)\theta_k^{\mathcal{B}}, \quad \mathbb{E}(d_k^{in,0}(X_k)) = (1-r)(1-\theta_k^{\mathcal{B}})$$
which leads to  $$  \alpha \big(d_k^{\text{in}, 1}(X_k) - d_k^{\text{in}, 0}(X_k)\big) - \beta \big(d_k^{\text{out}, 1}(X_k) - d_k^{\text{out}, 0}(X_k)\big) $$
$$ = \alpha((1-r)\theta_k^{\mathcal{B}}-(1-r)(1-\theta_k^{\mathcal{B}}))- \beta (r\theta_k^{\mathcal{R}}-r(1-\theta_k^{\mathcal{R}}))$$
$$ = \alpha(1 - r)(2\theta^\mathcal{B} - 1) - \beta r(2\theta^\mathcal{R} - 1).$$
Then, replacing the quantities in the decision rule in Eq.~\ref{eq:decision_rule} by their expected values yield the probabilities that a random blue node with choice-0 switches to 1 ($p_{\theta}^\mathcal{B}(0 \to 1)$) and vice-versa~($p_{\theta}^\mathcal{B}(1 \to 0)$). Based on those, $\dot{\theta}^\mathcal{B}$ yields the relative rate of change of ${\theta}^\mathcal{B}$. The corresponding expressions for the red-group follow from similar arguments. For $\delta = 0$ and $\theta^\mathcal{B}(0) = \theta^\mathcal{R}(0)$, the 
system \ref{eq:systemFC} has four possible equilibria: consensus where all individuals eventually adopt the same choice (case 1), partisan polarization where the choices are split along party lines (case 2 and 3) and non-partisan polarization where each group is split equally between the two choices (case 4).

\cite{nettasinghe2025outgroup} obtained several insights from the behavior of the model for a fully connected network. First, polarization requires the existence of out-group hate. Additionally, a high level of out-group hate compared to in-group love is enough to cause polarization. This aligns with the intuition that, in the case where people are more inclined to oppose to the out-group than to support the in-group, consensus would not occur. In particular, $\alpha < \beta$ leads to partisan or non-partisan polarization (\textbf{Case 2, 3, 4}). Moreover, a high level of in-group love compared to out-group hate is enough for consensus, provided that the group size imbalance is not too large. Finally, the majority cannot entirely switch to the most initially unpopular opinion. For example: if $r>0.5$ (i.e, the red group is the majority) then the system would fall into \textbf{Case 2} or \textbf{Case 4}. For both cases, the red group does not adopt the less popular choice. 

\vspace{0.1cm}
\noindent
{\bf Dynamics under a stochastic block model (SBM): } Let $G$ be sampled from a stochastic block model with two communities. An individual is connected to another node within the in-group and out-group with probabilities $\rho$ and $1- \rho$, respectively. The constant parameter $\rho \in (0,1)$ measures the level of homophily within the social network. For the stochastic block model, the mean field approximation to a continuous-time trajectory for large $N$ from Eqn.\ref{eq:systemFC} still holds, with certain modifications: $\alpha$ and $\beta $ are re-scaled by $\rho \alpha$ and $(1-\rho)\beta$, respectively. For $\delta = 0$ and $\theta^\mathcal{B}(0) = \theta^\mathcal{R}(0)$, the stochastic block model system behaves similarly as the fully connected model, with the respective re-scaling.

The main insights for the stochastic block model are as follows: when homophily is neutral (i.e, $\rho = 0.5$), it leads to an identical behavior of the stochastic block model system as a fully connected network. When $\rho = 0.5$, there is an identical probability of being connected to any node, independently of the group. Therefore, the structure of the network behaves as an Erd\H{o}s--R\'enyi model, which dynamics follow the pattern of a fully connected model, according to \cite{nettasinghe2025outgroup}. Moreover, emphasizing the decisions of the out-group  may induce polarization. The intuition is that, when $\rho < 0.5$ and $1-\rho > 0.5$, out-group hate might be amplified by emphasizing the out-group decisions.

Although the analysis of the model for fully connected networks and stochastic block models yield interesting results, many properties of real-world networks are not observed by those network models. For example shorter average path-lengths with smaller density \cite{doi:10.1073/pnas.98.2.404}, \cite{smallworldart}
and scale free degree distributions \cite{HJeong_2003}, \cite{liljeros2001sexconta}. Our results provided in the next two subsections aim to address these gaps.  

\section{Dynamics of Affective Polarization in Watts-Strogatz Networks}
\label{sec:SWModel}
\onehalfspacing

We first extend the model in \cite{nettasinghe2025outgroup} to a social network $G = (V,E)$ where $G$ is sampled from an augmented Watts-Strogatz model. The original Watts-Strogatz model is not bi-populated, i.e., has only a homogeneous group of nodes. Hence, to consider a binary attribute that represents the political affiliation of each node, an extension of the original Watts-Strogatz model is required.

\vspace{0.1cm}
\noindent
{\bf Augmented Watts-Strogatz Model:} The construction of this network is similar to the Watts-Strogatz Model given in~\cite{watts1998small}. We start from a ring lattice with nodes $1, 2,\dots, N$ and $d$ edges per vertex, where each node is initially connected to its $\frac{d}{2}$ nearest neighbors on each side. Subsequently, each edge is randomly rewired with probability $p$ \cite{watts1998small}. We then assign two binary attributes to each individual $v \in V$: node color (red or blue) and the initial choice (1 or 0). The nodes $1, 2, \dots, \lfloor Nr \rfloor$ are assigned the color red and the remaining nodes are assigned the color blue.\footnote{If one wishes to study the case in which the $\lfloor Nr \rfloor$ nodes are randomly selected to be colored red and the rest of the nodes be colored blue, it is equivalent to executing our original coloring algorithm with $p=1$.} Then, the choice-1 is assigned to $ \lfloor \theta_0^{\mathcal{B}}(1-r)N \rfloor$ random blue nodes and $ \lfloor \theta_0^{\mathcal{R}}rN \rfloor $ random red nodes, where $\theta_0^{\mathcal{B}}$ and $\theta_0^{\mathcal{R}}$ are the pre-determined fraction of blue and red nodes with choice one at time $k = 0$. All other nodes are assigned the initial choice-0.

For the above network model, the piece-wise interpolation of the discrete-time trajectory ${\theta}_k = [\theta_k^\mathcal{B}, \theta_k^\mathcal{R}]^\prime, k = 0,1,2,...$ can be approximated by a continuous-time trajectory in a manner similar to the fully connected and stochastic block models considered in \cite{nettasinghe2025outgroup}. This is feasible since the decision rule in Eq.~\ref{eq:decision_rule} depends on the structure of the network $G$ only via quantities defined in Eq.~\ref{eq:degrees} which can be expressed in terms of ${\theta}_k = [\theta_k^\mathcal{B}, \theta_k^\mathcal{R}]^\prime$, the fraction of the red nodes~$r$ and the rewiring probability $p$ in the augemented Watts-Strogatz model.

\subsection{Continuous Time Approximation of the Dynamics in Watts-Strogatz Network}

When the number of nodes $N$ is large and the average degree $d$ satisfies $d\leq \sqrt{N} ln(N)$, the trajectory ${\theta}(t) = [\theta^\mathcal{B}(t), \theta^\mathcal{R}(t)]^\prime, t \geq 0$ for the Watts-Strogatz model can be approximated by the following differential equation: 
\begin{equation}
\begin{bmatrix}
\dot{\theta}^\mathcal{B} \\
\dot{\theta}^\mathcal{R}
\end{bmatrix}
=
\begin{bmatrix}
(1 - \theta^\mathcal{B})p_\theta^\mathcal{B}(0 \to 1) - \theta^\mathcal{B} p_\theta^\mathcal{B}(1 \to 0) \\
(1 - \theta^\mathcal{R})p_\theta^\mathcal{R}(0 \to 1) - \theta^\mathcal{R}p_\theta^\mathcal{R}(1 \to 0)
\end{bmatrix}, 
\label{eq:systemSW}
\end{equation}
where,
\begin{equation*}
\begin{aligned}
p_{\theta}^\mathcal{B}(0 \to 1) &= \mathbbm{1} \left( \alpha(1 - pr)(2\theta^\mathcal{B} - 1) - \beta pr(2\theta^\mathcal{R} - 1) > \delta \right), \\
p_{\theta}^\mathcal{B}(1 \to 0) &= \mathbbm{1} \left( \alpha(1 - pr)(2\theta^\mathcal{B} - 1) - \beta pr(2\theta^\mathcal{R} - 1) < -\delta \right), \\
p_{\theta}^\mathcal{R}(0 \to 1) &= \mathbbm{1} \left( \alpha (1-p(1-r))(2\theta^\mathcal{R} - 1) - \beta(1 - r)p(2\theta^\mathcal{B} - 1) > \delta \right), \\
p_{\theta}^\mathcal{R}(1 \to 0) &= \mathbbm{1} \left( \alpha (1-p(1-r))(2\theta^\mathcal{R} - 1) - \beta(1 - r)p(2\theta^\mathcal{B} - 1) < -\delta \right).
\end{aligned}    
\end{equation*}

The intuition behind the above mean-field approximation is as follows. If we picked a blue node randomly with choice-0 at time $k$ then:  
 \begin{itemize}
        \item A fraction $(1-p)$ of neighbors are still connected from the ring lattice arrangement. In this arrangement, most of the blue nodes do not have red neighbors due to the smaller average degree (See Appendix~\ref{subsec:SIdmin} for details). 
        \item A fraction $p$ of neighbors is randomly rewired.
    \end{itemize}
Therefore, if the randomly picked node $X_k$  is blue, we have:
$$ \mathbb{E}(d_k^{out,1}(X_k)) = \frac{dpr\theta_k^{\mathcal{R}}}{d} = pr\theta_k^{\mathcal{R}}$$
$$ \mathbb{E}(d_k^{out,0}(X_k)) = \frac{dpr(1-\theta_k^{\mathcal{R}})}{d} = pr(1-\theta_k^{\mathcal{R}})$$
$$ \mathbb{E}(d_k^{in,1}(X_k)) = \frac{d(1-p)\theta_k^{\mathcal{B}}+dp(1-r)\theta_k^{\mathcal{B}}}{d} = (1-pr)\theta_k^{\mathcal{B}}$$
$$ \mathbb{E}(d_k^{in,0}(X_k)) = \frac{d(1-p)(1-\theta_k^{\mathcal{B}})+dp(1-r)(1-\theta_k^{\mathcal{B}})}{d} = (1-pr)(1-\theta_k^{\mathcal{B}})$$

Substituting the above in the decision rule in Eq.\ref{eq:decision_rule} yields:
\begin{equation*}
    \mathbb{E}[\alpha \big(d_k^{\text{in}, 1}(X_k) - d_k^{\text{in}, 0}(X_k)\big) - \beta \big(d_k^{\text{out}, 1}(X_k) - d_k^{\text{out}, 0}(X_k)\big)] = \alpha(1 - pr)(2\theta^\mathcal{B} - 1) - \beta pr(2\theta^\mathcal{R} - 1)
\end{equation*} which leads to the closed-form of $p_{\theta}^\mathcal{B}(0 \to 1)$ and $p_{\theta}^\mathcal{B}(1 \to 0)$. The probabilities $p_{\theta}^\mathcal{R}(0 \to 1)$ and $p_{\theta}^\mathcal{R}(1 \to 0)$ follow from similar arguments. 

When $p = 1$, the dynamics of choices specified above takes the same form as the dynamics on a fully connected graph (or an Erd\H{o}s--R\'enyi model) that we discussed in Sec.~\ref{sec:related_work}.

\subsection{Analysis of the Dynamics on Watts-Strogatz Model}

The theorem below characterizes the dynamics of the choices on the augmented Watts-Strogatz network. 

\begin{theorem}
     Consider Eq.~\ref{eq:systemSW}, which represents the dynamics of the state of the population $\theta(t) = [\theta^\mathcal{B}(t), \theta^\mathcal{R}(t)]^{'}$ under the affective polarization model in Eq.~\ref{eq:decision_rule} on a Watts-Strogatz graph with rewiring constant $p$. Let $\delta = 0$ (i.e. no inertia) and $\theta^\mathcal{B}(0) = \theta^\mathcal{R}(0)$ (i.e. initial state is party independent).  Then, the following statements characterize the asymptotic state of the system for various different values of $\alpha$ (level of in-group love), $\beta$ (level of out-group hate), $r$ (fraction of red nodes in the network) and $p$ (rewiring constant of the graph):
     \begin{itemize}
         \item Case 1: Let $ p < \frac{\alpha}{(\alpha+\beta) \cdot \max\{r,1-r\}}$. If $\theta^\mathcal{B}(0) = \theta^\mathcal{R}(0) > 0.5$, then $\lim_{t \to \infty} \theta(t) = [1,1]^{'}$. If $\theta^\mathcal{B}(0) = \theta^\mathcal{R}(0) < 0.5$, then $\lim_{t \to \infty} \theta(t)= [0,0]^{'}$ i.e., consensus will emerge with both groups fully adopting the choice that was initially more popular.  
         \item Case 2: Let $\frac{\alpha}{(\alpha+\beta) \cdot r} < p < \frac{\alpha}{(\alpha+\beta) \cdot (1-r)}$. If $\theta^\mathcal{B}(0) = \theta^\mathcal{R}(0) > 0.5$, then $\lim_{t \to \infty} \theta(t) = [0,1]^{'}$. If $\theta^\mathcal{B}(0) = \theta^\mathcal{R}(0) < 0.5$, then $\lim_{t \to \infty} \theta(t) = [1,0]^{'}$ i.e., party-line polarization will emerge with the red group (majority) fully adopting the choice that was initially more popular and the blue group fully adopting the other choice.
         \item Case 3: Let  $\frac{\alpha}{(\alpha+\beta) \cdot (1-r)} < p < \frac{\alpha}{(\alpha+\beta) \cdot r}$. If $\theta^\mathcal{B}(0) = \theta^\mathcal{R}(0) > 0.5$, then $\lim_{t \to \infty} \theta(t) = [1,0]^{'}$. If $\theta^\mathcal{B}(0) = \theta^\mathcal{R}(0) < 0.5$, then $\lim_{t \to \infty} \theta(t) = [0,1]^{'}$ i.e., party-line polarization will emerge with the blue group (majority) fully adopting the choice that was initially more popular and the red group fully adopting the other choice.
         \item Case 4: Let $ p > \frac{\alpha}{(\alpha+\beta) \cdot \min\{r,1-r\}}$. If $\theta^\mathcal{B}(0) = \theta^\mathcal{R}(0)$, then $\lim_{t \to \infty} \theta(t) = [0.5, 0.5]^{'}$  i.e., nonpartisan polarization will emerge with half of each group adopting choice-1 and the remaining half adopting choice-0.
     \end{itemize}
     \label{TheoremSW}
     \end{theorem}

\begin{figure}[h!]
    \centering

    \begin{minipage}[t]{0.48\textwidth}
        \centering
        \includegraphics[width=\linewidth]{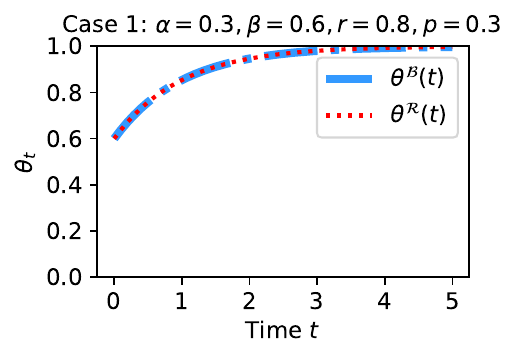}
    \end{minipage}
    \hfill
    \begin{minipage}[t]{0.48\textwidth}
        \centering
        \includegraphics[width=\linewidth]{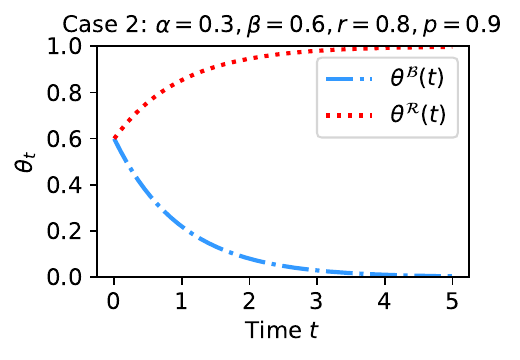}
    \end{minipage}


    \begin{minipage}[t]{0.48\textwidth}
        \centering
        \includegraphics[width=\linewidth]{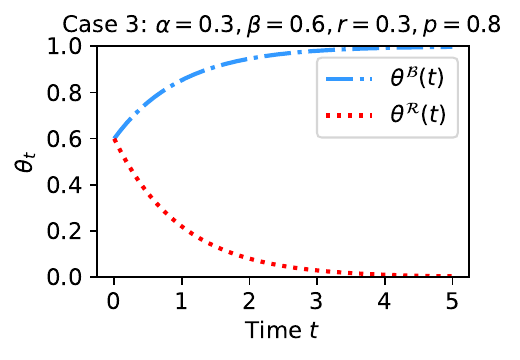}
    \end{minipage}
    \hfill
    \begin{minipage}[t]{0.48\textwidth}
        \centering
        \includegraphics[width=\linewidth]{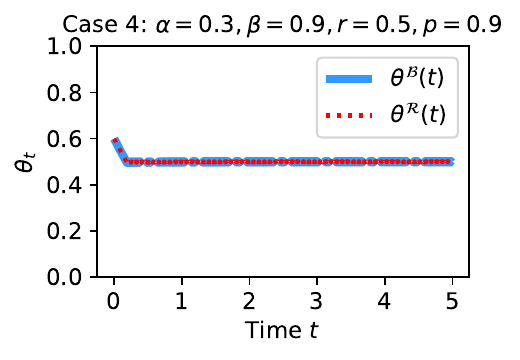}
    \end{minipage}

    \caption{Four example trajectories illustrating each case of {Theorem }\ref{TheoremSW} for $\theta^\mathcal{B}(0) = \theta^\mathcal{R}(0) = 0.6$. The long-term outcomes are: (case 1) No Polarization, (case 2/case 3) Partisan Polarization, and (case 4) Non-Partisan Polarization.}
    \label{fig:fourcasesThe1}
\end{figure}

Theorem~\ref{TheoremSW} shows the outcomes that can emerge in the Augmented Watts-Strogatz model under affective polarization: 
global consensus~(case 1), party line
polarization (case 2 and 3), and nonpartisan polarization~(case 4). The limiting states in cases 1–3 of Theorem~\ref{TheoremSW} (consensus and polarization along party lines) are locally asymptotically stable stationary states of the system in Eq.~\ref{eq:systemSW} whereas the limiting state in case~4 is an unstable stationary state of Eq. \ref{eq:systemSW}. Fig.~\ref{fig:fourcasesThe1} shows example trajectories for each of the four cases in Theorem~\ref{TheoremSW}. The proof of this theorem is provided in the Supplementary Information \ref{subsec:proofth1}.

 Next, we discuss several insights from Theorem \ref{TheoremSW}:
\vspace{0.1cm}

\noindent
\textbf{Within a ring-lattice society, consensus is ultimately guaranteed for small enough average degree:} A ring-lattice arrangement is a particular case of the Watts-Strogatz model with $p=0$, which according to Theorem \ref{TheoremSW}, will always fall into \textit{Case 1} for average degree $d \leq \sqrt{N} ln(N)$. The key reason is that people are only locally connected so that there are relatively fewer inter-group (between red and blue) ties, making the effect of out-group hate smaller. This insight aligns with findings from \cite{nettasinghe2025outgroup} which showed that less homophily in the stochastic block model can lead to polarization. \cite{polbackfire} also finds that nonlocal interactions via digital media contribute to the alignment of conflicts along partisan lines, leading to partisan polarization.

\vspace{0.1cm}
\noindent
\textbf{No in-group love leads to nonpartisan polarization~i.e.,~in-group love is indispensable for consensus:} According to Theorem \ref{TheoremSW}, if the in-group love $\alpha$ is closer to zero, then the system will necessarily behave as in case 4, which leads to non-partisan polarization from any party-independent initial state $\theta^\mathcal{B}(0) = \theta^\mathcal{R}(0) \neq 0.5$, making it impossible to reach consensus. This behavior is similar to the dynamics observed in fully connected networks and stochastic block model. 

\begin{figure}[H]
    \centering
    
    \begin{minipage}[t]{0.45\textwidth}
        \centering
        \includegraphics[width=\linewidth]{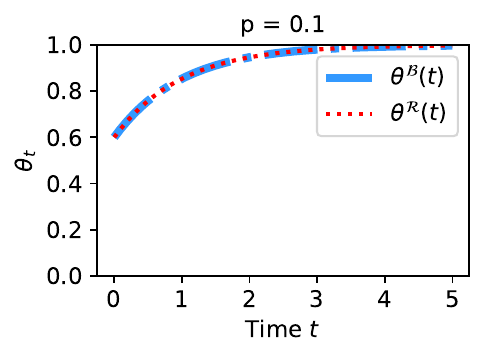}
        
    \end{minipage}
    \begin{minipage}[t]{0.45\textwidth}
        \centering
        \includegraphics[width=\linewidth]{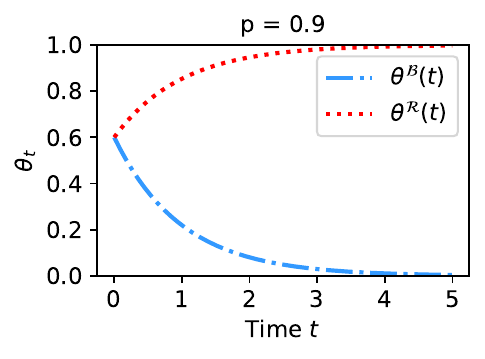}
        
    \end{minipage}

    \caption{Two example trajectories illustrating how a smaller rewiring constant $p$ facilitates reaching consensus. In all trajectories $\theta^\mathcal{B}(0) = \theta^\mathcal{R}(0) = 0.6, r = 0.8, \alpha = 0.4, \beta = 0.2$.}
    \label{fig:threeins}
\end{figure}

\vspace{0.1cm}
\noindent
\textbf{Larger in-group love relative to out-group hate leads to consensus as long as the rewiring constant is not too large :}
When the groups have the same size ($r = 0.5$), as long as the rewiring constant satisfies $p \le 0.5$, larger in-group love relative to out-group hate (i.e. $\beta < \alpha$) assures that all the individuals will adopt the most popular choice, leading to consensus. Moreover, for different group sizes, the same behavior holds. Making larger in-group love relative to out-group hate and a small rewiring constant ($p \le 0.5$) are sufficient conditions to reach consensus, as shown in Fig.\ref{fig:threeins}. Unlike the dynamics in fully connected networks summarized in Sec.~\ref{sec:related_work} where consensus requires a small group imbalance when in-group love is larger than out-group hate, the Augmented Watts-Strogatz model reaches consensus with only a small rewiring constant, allowing consensus even for significantly different group proportions. 
\vspace{0.1cm}

\noindent
\textbf{A smaller rewiring constant $p$ facilitates reaching consensus:} The idea behind this phenomenon is the following: people related to a higher proportion of their in-group are more likely to reach consensus than those related to a higher proportion of their out-group.

\noindent
\vspace{0.1cm}
\textbf{Similar to the fully connected model and stochastic block model summarized in Sec.~\ref{sec:related_work}, majority cannot fully adopt the initially most unpopular choice:} Without loss of generality, let $r > 0.5$ and $\theta^\mathcal{B}(0) = \theta^\mathcal{R}(0) > 0.5$. The only case where the red group could fully adopt choice-0 would be falling into case 3 of Theorem \ref{TheoremSW}. However, as $ \frac{\alpha}{(\alpha+\beta) \cdot r}  < \frac{\alpha}{(\alpha+\beta) \cdot (1-r)} $ for $r>0.5$, this case never occurs. The case in which the majority is the blue group is similar.

Overall, the Watts-Strogatz model provides additional insights beyond those present in the fully connected and stochastic block model dynamics. In particular, a smaller average path length of
the network (resembling a small-world network) makes achieving the consensus further difficult (since a smaller rewiring constant p facilitates reaching consensus). The analysis of the dynamics in more complex network structures allowed us to examine the role of structural properties such as the average path length.   

\section{Dynamics of Affective Polarization in Configuration Model}
\label{sec:ConfModel}
Next, we consider a social network $G = (V,E)$ sampled from the Configuration Model \cite{CMMolloy} with a given degree distribution for the blue and red group. Compared to previous cases discussed so far, dynamics on such networks characterized by their degree distributions require a different approach. To understand the main challenge, let us consider the model in Eq.~\ref{eq:decision_rule} again. In all the previous cases (fully connected, stochastic block model, augmented watts-strogatz), we used ${\theta}_k = [\theta_k^\mathcal{B}, \theta_k^\mathcal{R}]^\prime$, which is the fraction of nodes with choice-1 in each group, as the state of the system, and then expressed it as a recursion. This was possible because, conditional on the group (color) and the current choice (1 or 0), the choice of the next time instant is independent of everything else except ${\theta}_k = [\theta_k^\mathcal{B}, \theta_k^\mathcal{R}]^\prime$ in those cases. However, for networks with arbitrary degree distributions, two nodes from the same group with the same current choice may behave differently depending on their number of in-group and out-group connections.

To deal with this context, we define the state at time $k$ to be the fraction of nodes in each group (blue, red) with $f$ friends (same group connections) and $e$ enemies (opposite group connections), that have adopted choice-1 at time \( k \). Formally, for each $f,e \in \{0,1,...,N-1\}$, we define the state of the system at time \( k \) as the column vector ${\theta}_{k,f,e} = [\theta_{k,f,e}^\mathcal{B}, \theta_{k,f,e}^\mathcal{R}]^\prime $ where,
\begin{equation}
\theta_{k,f,e}^\mathcal{B} = \frac{\sum_{v \in V} \mathbbm{1}(R(v) = 0 \land H_k(v) = 1 \land d^{in}(v) = f \land d^{out}(v) = e)}{\sum_{v \in V} \mathbbm{1}(R(v) = 0  \land d^{in}(v) = f \land d^{out}(v) = e)}
    \label{eqn:thetageneral}
\end{equation}
\begin{equation*}
    \theta_{k,f,e}^\mathcal{R} = \frac{\sum_{v \in V} \mathbbm{1}(R(v) = 1 \land H_k(v) = 1 \land d^{in}(v) = f \land d^{out}(v) = e)}{\sum_{v \in V} \mathbbm{1}(R(v) = 1  \land d^{in}(v) = f \land d^{out}(v) = e)}.
\end{equation*}
where $ d^{in}(v) = d^{in,1}(v)+d^{in,0}(v)$ and $ d^{out}(v) = d^{out,1}(v)+d^{out,0}(v)$ (based on notation defined in Sec.\ref{sec:related_work}, Eq.~\ref{eq:degrees}).

Thus, the dynamics in the configuration model are represented with a state that can be much higher in dimension compared to the previous cases, and consequently more challenging to be analyzed.

\subsection{Continuous Time Approximation of the Dynamics in a Configuration Network}
 Let $P_{f,e}^\mathcal{R}$ (resp. $P_{f,e}^\mathcal{B}$) denote the degree distribution of the red~(resp.~blue) group~i.e.,~$P_{f,e}^\mathcal{R}$ (resp. $P_{f,e}^\mathcal{B}$) gives the probability that a random red (resp.~blue) node has $f$ friends and $e$ enemies. The piece-wise interpolation of the discrete-time trajectory of the state ${\theta}_{k,f,e} = [\theta_{k,f,e}^\mathcal{B}, \theta_{k,f,e}^\mathcal{R}]^\prime, k = 0,1,2,...$ and $ f,e \in \{0,...,N-1 \}$ can be approximated using the continuous-time trajectory  \\ ${\theta_{f,e}}(t) = [\theta_{f,e}^\mathcal{B}(t), \theta_{f,e}^\mathcal{R}(t)]^\prime, t \geq 0 $ of the following differential equation as the number of nodes $N$ is large: 
\begin{equation}
\begin{bmatrix}
\dot{\theta}_{f,e}^\mathcal{B} \\
\dot{\theta}_{f,e}^\mathcal{R}
\end{bmatrix}
=
\begin{bmatrix}
(1 - \theta_{f,e}^\mathcal{B})p_{f,e}^\mathcal{B}(0 \to 1) - \theta_{f,e}^\mathcal{B} p_{f,e}^\mathcal{B}(1 \to 0) \\
(1 - \theta_{f,e}^\mathcal{R})p_{f,e}^\mathcal{R}(0 \to 1) - \theta_{f,e}^\mathcal{R} p_{f,e}^\mathcal{R}(1 \to 0)
\end{bmatrix}, 
\label{eq:systemgeneralized}
\end{equation}

where,
\begin{equation*}
\begin{aligned}
p_{f,e}^\mathcal{B}(0 \to 1) &= \mathbbm{1} \left( \alpha \frac{f}{f+e}(2P_F^\mathcal{B}(t) - 1) - \beta \frac{e}{f+e}(2P_E^\mathcal{B}(t) - 1) > \delta \right), \\
p_{f,e}^\mathcal{B}(1 \to 0) &= \mathbbm{1} \left( \alpha \frac{f}{f+e}(2P_F^\mathcal{B}(t) - 1) - \beta \frac{e}{f+e}(2P_E^\mathcal{B}(t) - 1)< -\delta \right), \\
p_{f,e}^\mathcal{R}(0 \to 1) &= \mathbbm{1} \left( \alpha \frac{f}{f+e}(2P_F^\mathcal{R}(t) - 1) - \beta \frac{e}{f+e}(2P_E^\mathcal{R}(t) - 1)> \delta \right), \\
p_{f,e}^\mathcal{R}(1 \to 0) &= \mathbbm{1} \left(\alpha \frac{f}{f+e}(2P_F^\mathcal{R}(t) - 1) - \beta \frac{e}{f+e}(2P_E^\mathcal{R}(t) - 1)< -\delta \right),
\end{aligned}    
\end{equation*}
and,
 \begin{align}
       P_F^\mathcal{R}(t) = \frac{\sum_{e, f} f P_{f,e}^\mathcal{R} \theta_{f,e}^\mathcal{R}(t)}{\sum_{e, f} f P_{f,e}^\mathcal{R}},  P_E^\mathcal{R}(t) = \frac{\sum_{e, f} e P_{f,e}^\mathcal{B} \theta_{f,e}^\mathcal{B}(t)}{\sum_{e, f} e P_{f,e}^\mathcal{B}} \\
       P_F^\mathcal{B}(t) = \frac{\sum_{e, f} f P_{f,e}^\mathcal{B} \theta_{f,e}^\mathcal{B}(t)}{\sum_{e, f} f P_{f,e}^\mathcal{B}},  P_E^\mathcal{B} = \frac{\sum_{e, f} e P_{f,e}^\mathcal{R}(t) \theta_{f,e}^\mathcal{R}(t)}{\sum_{e, f} e P_{f,e}^\mathcal{R}}.
     \label{eqn:probs_b_r}
 \end{align}

In the above mean-field model, $P_F^\mathcal{R}(t)$ and $P_E^\mathcal{R}(t)$ indicate the probabilities that a random friend of a red node has choice-1 and the probability that a random enemy of a red node has choice-1, respectively, at time $t$. Analogous quantities for the blue group are denoted by $P_F^\mathcal{B}(t)$ and $P_E^\mathcal{B}(t)$. To understand the intuition, consider picking a random friend of a random red node. In other words, from all links that go between the red nodes, we pick one randomly and then take one end of that link with a coin toss. A red node with $f$ number of friends is $f$ times more likely to be picked in this process than a red node with $1$ friend. Thus, $fP_{f,e}^\mathcal{R}/\sum_{e,f}fP_{f,e}^\mathcal{R} $ is the probability of sampling a friend of a red node who has $f$ friends and $e$ enemies. Consequently, $fP_{f,e}^\mathcal{R}\theta_{f,e}^\mathcal{R}(t)/\sum_{e,f}fP_{f,e}^\mathcal{R}$ is the probability of sampling a random friend of a red node with $f$ friends, $e$ enemies and has choice-1, yielding the expression for $P_F^\mathcal{R}(t)$. Similar arguments yield expressions for the $P_E^\mathcal{R}(t), P_F^\mathcal{B}(t)$ and $P_E^\mathcal{B}(t)$ as well. Then, for a blue node with $f$ friends and $e$ enemies, we can write,
$$ \mathbb{E}(d_k^{in,1}(X_k)) = \frac{fP_F^\mathcal{B}(t)}{e+f}, \quad \mathbb{E}(d_k^{in,0}(X_k)) = \frac{f(1-P_F^\mathcal{B}(t))}{e+f}$$
$$ \mathbb{E}(d_k^{out,1}(X_k)) = \frac{eP_E^\mathcal{B}(t)}{e+f}, \quad \mathbb{E}(d_k^{out,0}(X_k)) = \frac{e(1-P_E^\mathcal{B}(t))}{e+f},$$
 and the mean-field model in Eq.~\ref{eq:systemgeneralized} follows from Eq.~\ref{eq:decision_rule}. When the network is fully connected, Eq.\ref{eq:systemgeneralized} reduces to the same expression in Eq.~\ref{eq:systemFC}.

For the rest of this paper, we assume that the friend and enemy degree distributions of each group have scale-free forms and they are independent of each other. More specifically, we assume that $P_{f,e}^\mathcal{R} \propto e^{-\gamma_\mathcal{R}}f^{-\eta_\mathcal{R}}$ and $P_{f,e}^\mathcal{B} \propto e^{-\gamma_\mathcal{B}}f^{-\eta_\mathcal{B}}$ where $\gamma_\mathcal{R},\gamma_\mathcal{B},\eta_\mathcal{R},\eta_\mathcal{B}$ denote the scale-free exponents of friend and enemy degree distributions of red and blue groups.\footnote{Preferential attachment models have recently generalized to bi-populated settings where the degree distribution of each group follow such power-law forms~\cite{nettasinghe2026emergence, nettasinghe2022scale}.} We also assume $\delta = 0$ in Eq.~\ref{eq:systemgeneralized} for the subsequent analysis.

\subsection{Comparison of the Mean-Field Approximation with Actual Dynamics}

Before proceeding to analyze the mean-field dynamics presented in Eq.~\ref{eq:systemgeneralized}, we illustrate the mean-field approximation aligns with the actual model dynamics via several numerical experiments.

\vspace{0.1cm}
\noindent
{\bf Experimental Setup: } We use the configuration model~\cite{confmodel}, \cite{CMMolloy} to generate networks with scale-free friend and enemy degree distributions given by $P_{f,e}^\mathcal{R} \propto e^{-\gamma_\mathcal{R}}f^{-\eta_\mathcal{R}}$ and $P_{f,e}^\mathcal{B} \propto e^{-\gamma_\mathcal{B}}f^{-\eta_\mathcal{B}}$ for $e,f$ in the range $d_{\text{min}} = 1, d_{\text{max}} = 100$ as follows. We first create $70000$ nodes for each group to get a total of $140000$ nodes. For each blue node, the number of friend half-edges (stubs) and enemy half-edges are sampled from the (unnormalized) probability distribution $f^{-\eta_\mathcal{B}}$ and the enemy degree is sampled from the probability distribution $e^{-\gamma_\mathcal{B}}$. The friend and enemy half-edges for the red-group are assigned similarly. Then, the half-edges are paired randomly ensuring enemy edges connecting the two groups and friend edges staying within the same group. We discard any self-loops or multi-edges and do repairing until a valid edge is created. For further details on the method used to construct the graphs, see Sec.~\ref{subsec:consCM}. Once the network is created, a fraction $\theta_{f,e}^\mathcal{B}(0)$ of all blue nodes with $f$ friends and $e$ enemies are assigned choice-1 at time $t = 0$ and the same process is followed for the red-group. Then, the model (described by Eq.~\ref{eq:decision_rule}) is executed for $K = 1000$ discrete time steps. We then visualize the fraction of nodes that have choice-1 at the last time instant for each $f,e$ bin in each group~i.e.,~$\theta_{k,f,e}^\mathcal{B}$ and $\theta_{k,f,e}^\mathcal{R}$ at time $k = K$ is plotted against $f,e$. We also obtain the mean-field approximation $\theta_{f,e}^\mathcal{B}(t), \theta_{f,e}^\mathcal{R}(t)$ via Eq.~\ref{eq:systemgeneralized}.

\begin{figure}[H]
\centering

\begin{subfigure}{0.48\textwidth}
    \centering
    \includegraphics[width=\linewidth]{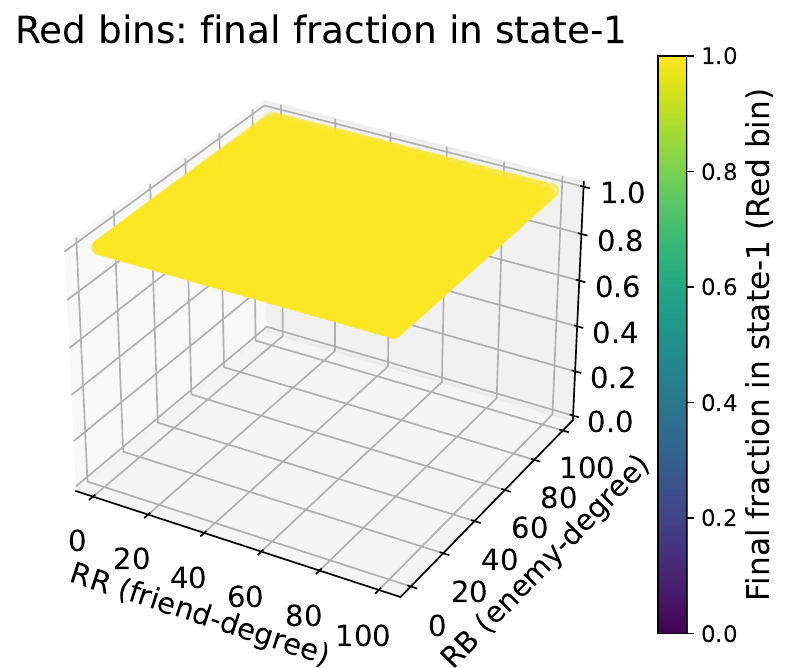}
    \caption{}
    \label{fig:redcont1}
\end{subfigure}
\hfill
\begin{subfigure}{0.48\textwidth}
    \centering
    \includegraphics[width=\linewidth]{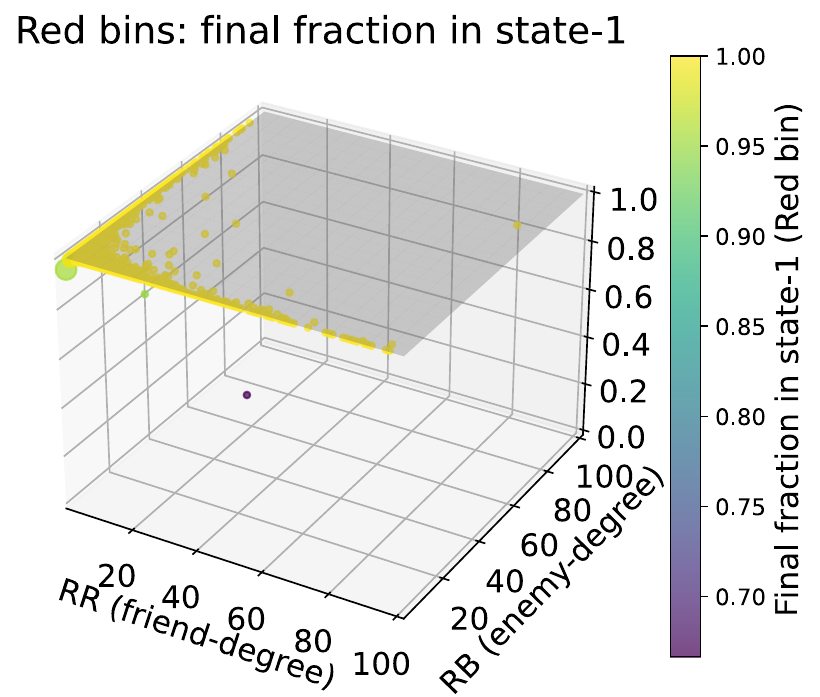}
    \caption{}
    \label{fig:reddis1}
\end{subfigure}

\vspace{0.3cm}

\begin{subfigure}{0.48\textwidth}
    \centering
    \includegraphics[width=\linewidth]{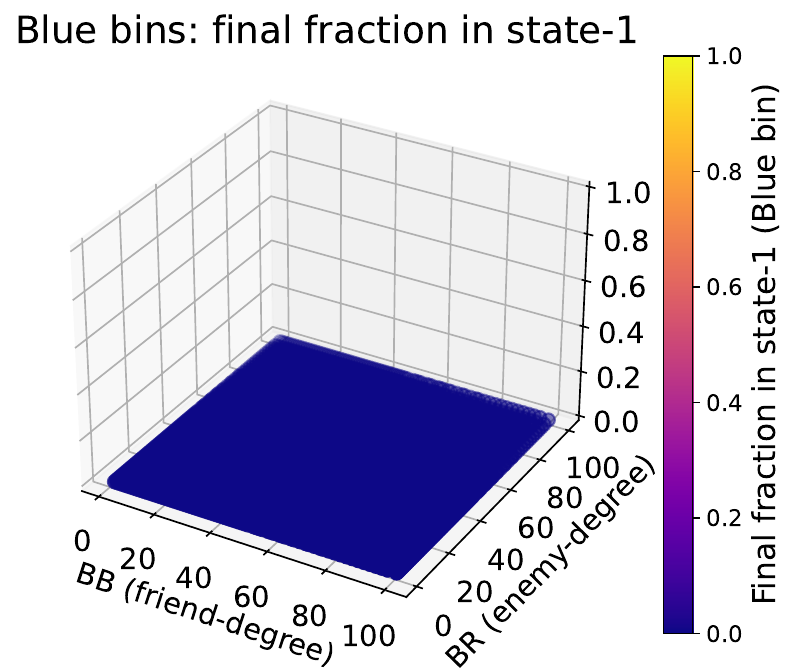}
    \caption{}
    \label{fig:bluecont1}
\end{subfigure}
\hfill
\begin{subfigure}{0.48\textwidth}
    \centering
    \includegraphics[width=\linewidth]{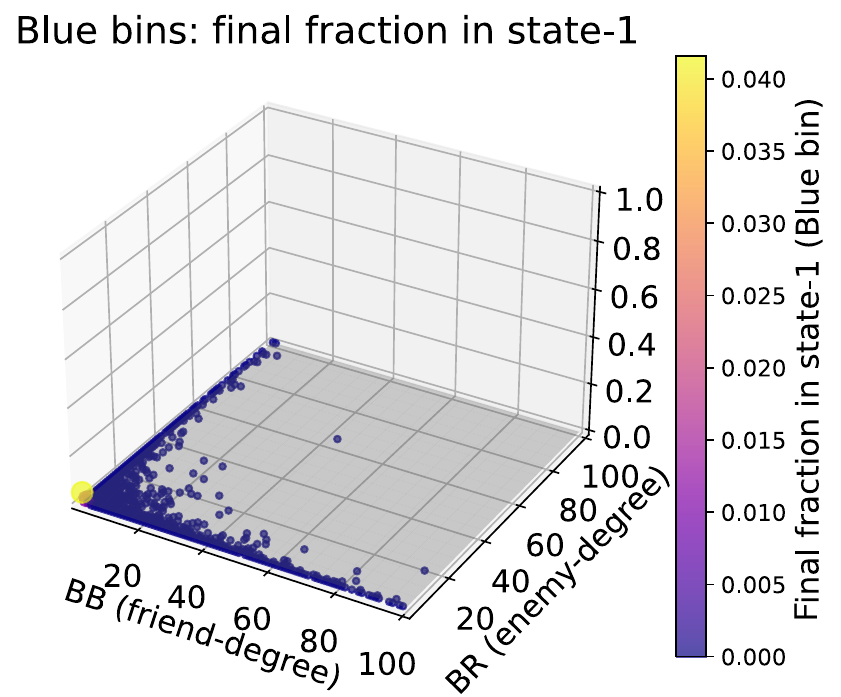}
    \caption{}
    \label{fig:bluedis1}
\end{subfigure}

\caption{Discrete simulation of the model (column-2) and mean-field approximation of the model (column-1). Fraction of people in the blue (row-2) and red (row-1) group of each bin that adopted choice-1 in the final state is plotted. For each case the following conditions were fixed: $d_{min} = 1, d_{max} = 100, \alpha = 0.6, \beta = 0.6, \eta_\mathcal{R} = 2.3, \eta_\mathcal{B} = 2.1, \gamma_\mathcal{R} = \gamma_\mathcal{B} = 2.4, \theta_{f,e}^\mathcal{B}(0) = 0.2, \theta_{f,e}^\mathcal{R}(0) = 0.8 $. This figure shows that the  mean-field approximation replicates the dynamics of the discrete model under the conditions above.}
\label{fig:ApproxPAModel1}

\end{figure}

\begin{figure}[H]
\centering

\begin{subfigure}{0.48\textwidth}
    \centering
    \includegraphics[width=\linewidth]{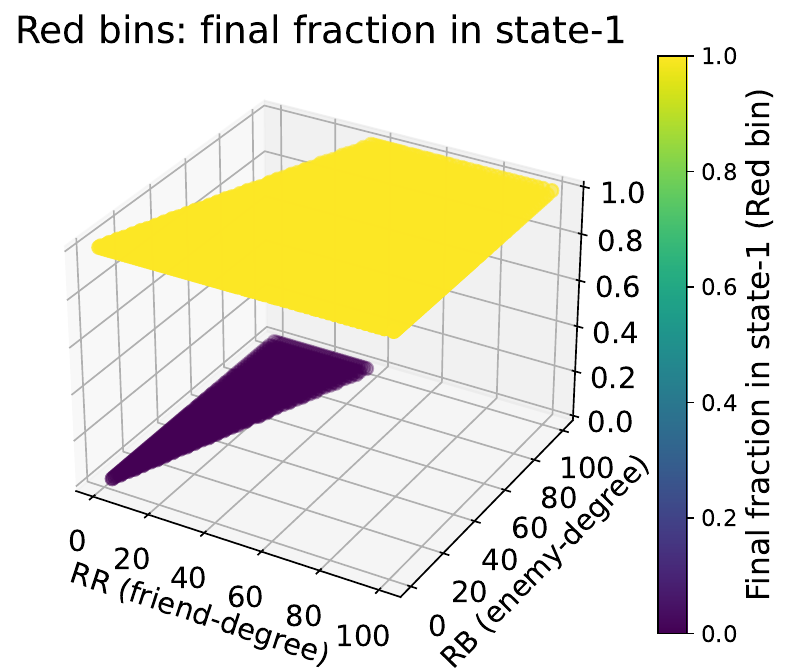}
    \caption{}
    \label{fig:redcont2}
\end{subfigure}
\hfill
\begin{subfigure}{0.48\textwidth}
    \centering
    \includegraphics[width=\linewidth]{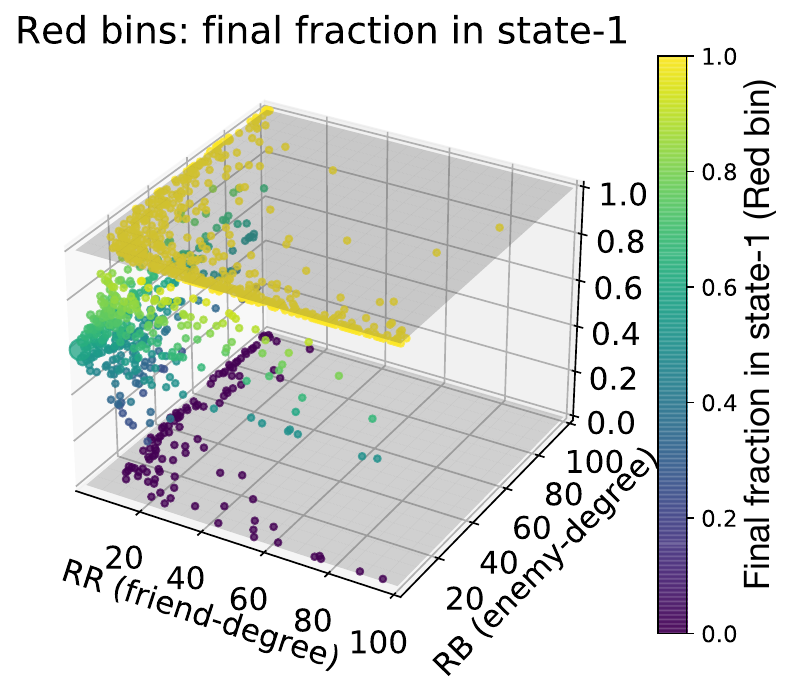}
    \caption{}
    \label{fig:reddis2}
\end{subfigure}

\vspace{0.3cm}

\begin{subfigure}{0.48\textwidth}
    \centering
    \includegraphics[width=\linewidth]{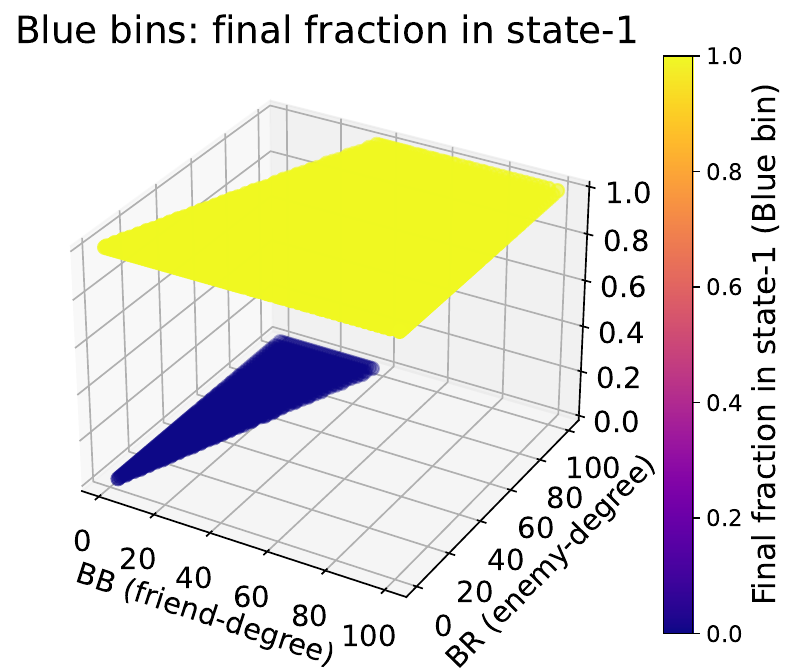}
    \caption{}
    \label{fig:bluecont2}
\end{subfigure}
\hfill
\begin{subfigure}{0.48\textwidth}
    \centering
    \includegraphics[width=\linewidth]{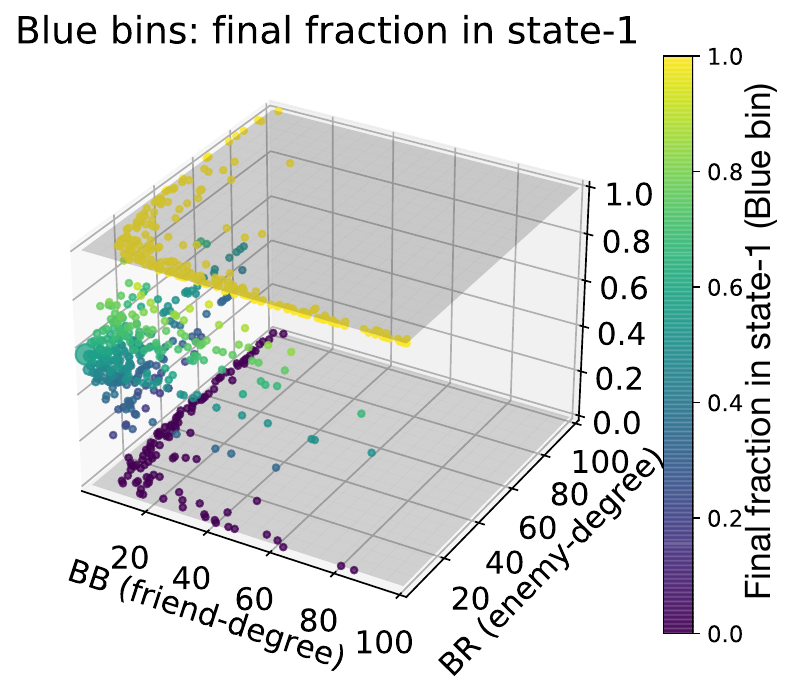}
    \caption{}
    \label{fig:bluedis2}
\end{subfigure}

\caption{Discrete simulation of the model (column-2) and mean-field approximation of the model (column-1). Fraction of people in the blue (row-2) and red (row-1) group of each bin that adopted choice-1 in the final state is plotted. For each case the following conditions were fixed: $d_{min} = 1, d_{max} = 100, \alpha = 0.6, \beta = 0.6, \eta_ \mathcal{R}= \eta_\mathcal{B} = 2.2, \gamma_\mathcal{R} = \gamma_\mathcal{B}= 2.4, \theta_{f,e}^\mathcal{B}(0) = \theta_{f,e}^\mathcal{R}(0) = 0.6 $. This figure shows that the  mean-field approximation is similar to the dynamics of the discrete model under the conditions above.}
\label{fig:ApproxPAModel2}

\end{figure}

\vspace{0.1cm}
\noindent
{\bf Comparison of the Model Dynamics and Mean-Field Dynamics: } 
Fig.~\ref{fig:ApproxPAModel1} and Fig.~\ref{fig:ApproxPAModel2} compare the state of the continuous time mean-field model in Eq.~\ref{eq:systemgeneralized} ($\theta_{f,e}^\mathcal{B}(t), \theta_{f,e}^\mathcal{R}(t)$) at time $t = 1000$ (column-1) with the final state $\theta_{K,f,e}^\mathcal{B},\theta_{K,f,e}^\mathcal{R}$ from the simulation of the model on the network from the configuration model (column-2) under two different parameter configurations. These yield several insights.

Fig.~\ref{fig:ApproxPAModel1} corresponds to the parameter configuration $\alpha =\beta = 0.6, \eta_\mathcal{R} = 2.3, \eta_\mathcal{B} = 2.1, \gamma_\mathcal{R} = \gamma_\mathcal{B} = 2.4, \theta_{f,e}^\mathcal{B}(0) = 0.2, \theta_{f,e}^\mathcal{R}(0) = 0.8$. The continuous time mean-field dynamics (Fig.~\ref{fig:redcont1} and Fig.~\ref{fig:bluecont1}) shows that all red nodes adopt choice-1 and all blue nodes adopt choice-0 at equilibrium, irrespective of the number of friends and enemies~i.e.,~$\theta_{f,e}^\mathcal{R}(t)$ converges to $1$ and $\theta_{f,e}^\mathcal{B}(t)$ converges to 0 for all $f,e$. The results of the actual model (Fig.~\ref{fig:reddis1} and Fig.~\ref{fig:bluedis1}) agree with the mean-field model: $\theta_{k,f,e}^\mathcal{B},\theta_{k,f,e}^\mathcal{R}$ shows convergence to 0 and 1, respectively. In other words, red nodes in each $f,e$ bin have adopted choice-1 and blue nodes in each $f,e$ bin have adopted choice-0. The reason for this behavior is that $80\%$ in the red group have choice-1 initially whereas only $20\%$ in the blue-group has choice-1. Further, nodes in each group tend to have more friends than enemies since friend distribution has a smaller exponent than enemy exponent. Consequently, each group ultimately fully adopts the choice that was initially popular. We note that, not all bins have non-zero number of nodes in the actual power-law graph (Fig.~\ref{fig:reddis1} and Fig.~\ref{fig:bluedis1}) . In particular, bins corresponding to higher $f,e$ values have no nodes. This is because node degrees are sampled from a power-law distribution, and a finite sample size is unlikely to yield samples from the tail of a power-law distribution.

Fig.~\ref{fig:ApproxPAModel2} corresponds to the parameter configuration $\alpha = 0.6, \beta = 0.6, \eta_ \mathcal{R}= \eta_\mathcal{B} = 2.2, \gamma_\mathcal{R} = \gamma_\mathcal{B}= 2.4, \theta_{f,e}^\mathcal{B}(0) = \theta_{f,e}^\mathcal{R}(0) = 0.6$. Therefore, the two groups have the same parameter configurations and initial conditions. The mean-field dynamics ((Fig.~\ref{fig:redcont2} and Fig.~\ref{fig:bluecont2})) show that each group gets split between the two choices. More specifically, the individuals with more enemies than friends (i.e.,~the nodes in bins $f,e$ such that $e\gg f$) adopt choice-0 while others adopt choice-1 (which is the most popular choice initially in each bin). This observation is intuitive: since nodes with more enemies than friends have a larger enemy influence compared to friends, they give up the initially more popular choice-1 to differ from their enemies. Others who have more friends adopt the initially more popular choice. In this case also, the agreement between the mean-field approximation and the actual simulation is visible but less so compared to Fig.~\ref{fig:ApproxPAModel1}. The bins at the margins that separate the choice-1 group from the choice-2 have not converged fully to 1 or 0, likely due to the finite number of time steps in the simulations and the finite size of the network.

This comparison shows that the mean field model in Eq.\ref{eq:systemgeneralized} can be utilized instead of the computationally expensive simulations of the model on networks. In next sections, we implement the model to theoretically and numerically understand the different behaviors that can emerge as a consequence of affective polarization on scale-free networks.

\subsection{Analysis of the Dynamics on Configuration Model }

In this section, we first explore the theoretical properties of the mean-field dynamics in Eq.~\ref{eq:systemgeneralized} under the configuration model. We also provide numerical results that complement the theoretical results.

\subsubsection{Theoretical Results on the Dynamics on Configuration Model}

The following proposition provides sufficient conditions for consensus and non-partisan polarization to be achieved under the dynamics described in Eq. \ref{eq:systemgeneralized}. 
\begin{proposition} 
    Consider system \ref{eq:systemgeneralized} which represents the dynamics of the state of the population  ${\theta}_{f,e}(t) = [\theta_{f,e}^\mathcal{B}(t), \theta_{f,e}^\mathcal{R}(t)]^\prime$ for $f,e = 1,...,N-1$ under the model in Eq.~\ref{eq:decision_rule} in a social network sampled from the Configuration Model. Let  $e_{max} = \max_{v \in V}d^{out}(v)$, $e_{min} = \min_{v \in V}d^{out}(v)$, $f_{max} = \max_{v \in V} d^{in}(v)$ and $f_{min} = \min_{v \in V} d^{in}(v)$. If $\delta = 0$ (i.e. no inertia), $\theta_{f,e}^\mathcal{B}(0) = \theta_{f,e}^\mathcal{R}(0) = c$ (i.e., initial state is party independent), then the following statements provide sufficient conditions for consensus, depending on $\alpha$ (level of in-group love) and $\beta$ (level of out-group hate):   
    \label{prop:PA}
    \begin{itemize}
        \item Case 1. Let $\frac{\alpha}{\beta} > \frac{e_{max}}{f_{min}}$. If $c>0.5$, then $\lim_{t \to \infty} \theta_{f,e}(t) = [1,1]^{'}$ for all $f,e$. If $c< 0.5$, then $\lim_{t \to \infty} \theta_{f,e}(t) = [0,0]^{'}$ for all $f,e$. 
        \item Case 2. Let $\frac{\alpha}{\beta} < \frac{e_{min}}{f_{max}} $, then $\lim_{t \to \infty} \theta_{f,e}(t) = [\frac{1}{2},\frac{1}{2}]^{'}$ for all $f,e$. 
    \end{itemize}
\end{proposition}

The proof of this proposition is given below in \ref{subsec:proofprop1}. The idea behind it is to consider the full system ${\theta}(t) = [\theta_{f,e}^\mathcal{B}(t), \theta_{f,e}^\mathcal{R}(t)]_{\{f,e=1,...,N-1\}}$ and show that the points $A = [1,1,1,...] $ and $ B = [0,0,0,...] $ are fixed points of the system if and only if  the ratio $\frac{\alpha}{\beta} $ is bounded as in case 1. 

The intuition behind Proposition \ref{prop:PA} is the following. The sufficient condition for consensus in case 1 is equivalent to $\alpha f_{min} > \beta e_{max}$. This states that even for individuals with few friends and many enemies (i.e., friends degree equal to $f_{min}$ and enemies degree equal to $e_{max}$), the out-group hate effect does not dominate over the in-group love effect, leading to all individuals adopting the initially more popular choice. However, in case 2, even for individuals with few enemies and many friends (i.e., friends degree equal to $f_{max}$ and enemies degree equal to $e_{min}$) out-group hate dominates over in-group love, leading to non partisan polarization.

\begin{figure}[h!]
    \centering

    \begin{minipage}[t]{0.48\textwidth}
        \centering
        \includegraphics[width=\linewidth]{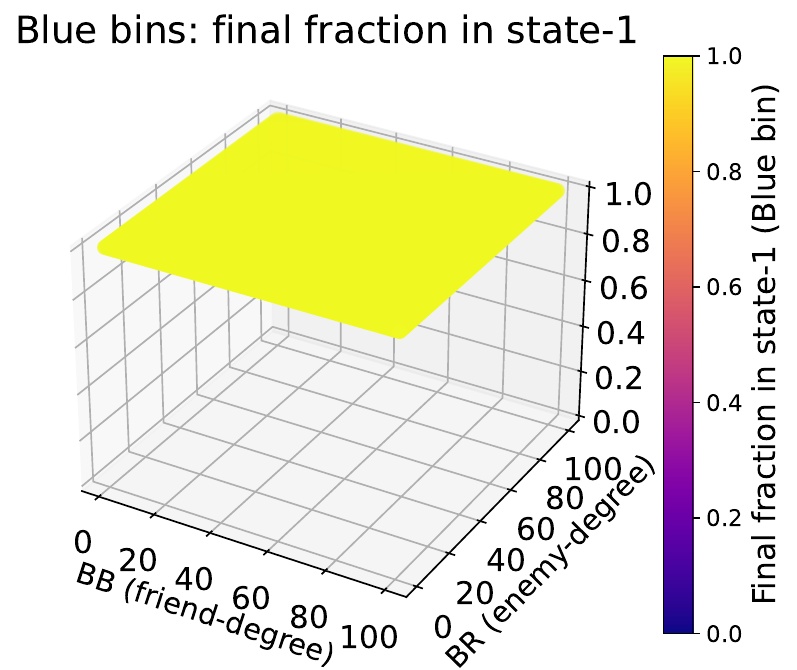}
        \caption*{(a)}
    \end{minipage}
    \hfill
    \begin{minipage}[t]{0.48\textwidth}
        \centering
        \includegraphics[width=\linewidth]{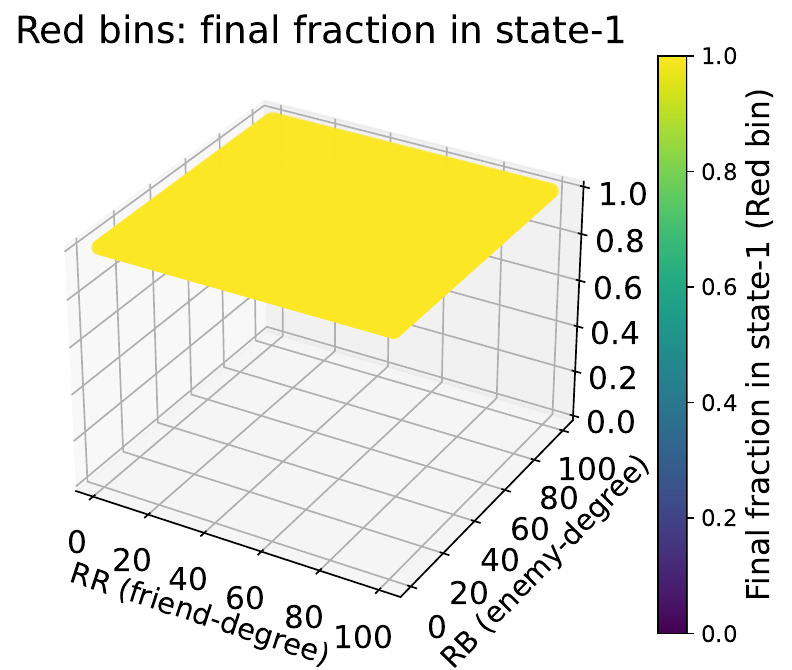}
        \caption*{(b)}
    \end{minipage}

    \vspace{0.3cm} 

    \begin{minipage}[t]{0.48\textwidth}
        \centering
        \includegraphics[width=\linewidth]{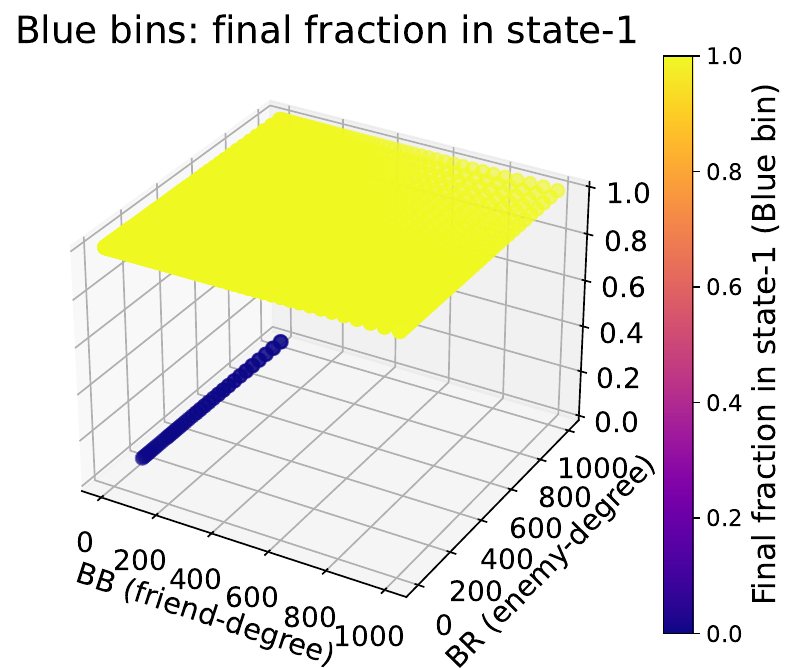}
        \caption*{(c)}
    \end{minipage}
    \hfill
    \begin{minipage}[t]{0.48\textwidth}
        \centering
        \includegraphics[width=\linewidth]{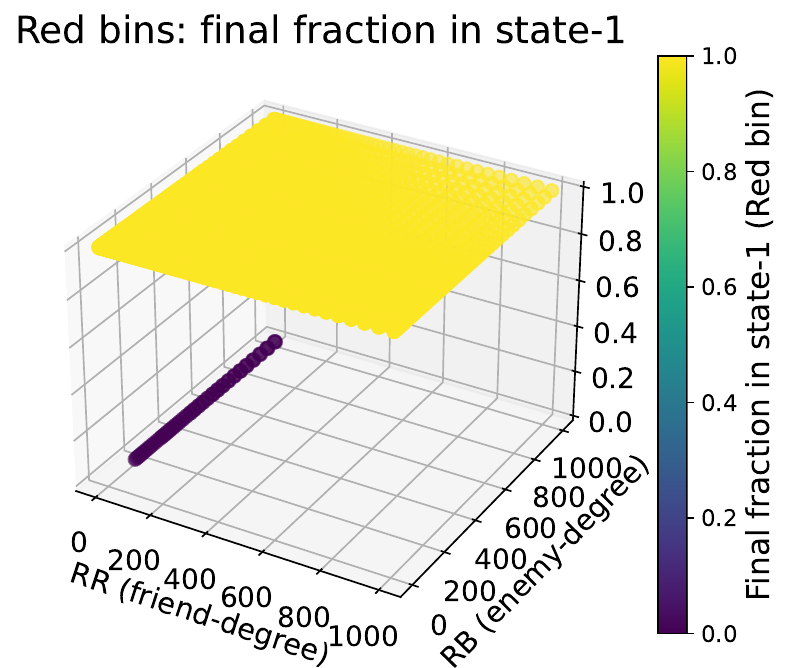}
        \caption*{(d)}
    \end{minipage}

    \caption{Fraction of people in the blue (column-1) and red (column-2) group of each bin that adopted choice-1 in the final state for $d_{max} = 100$ (row-1) and $d_{max} = 1000$ (row-2). For each case, the following conditions were fixed: $d_{min} = 1, \alpha = 0.15, \beta = 0.001, \eta_\mathcal{R} = 2.3, \eta_\mathcal{B} = 2.1, \gamma_\mathcal{R} = \gamma_\mathcal{B} = 2.4, \theta_{f,e}^\mathcal{B}(0) = \theta_{f,e}^\mathcal{R}(0) = 0.6 $. The behavior in both cases illustrates the results of Proposition \ref{prop:PA}.}
    \label{fig:FigPropPA1} 
\end{figure}

To illustrate Proposition \ref{prop:PA} numerically, we consider the mean field approximation model under the fixed parameters $d_{min} = 1, \alpha = 0.15, \beta = 0.001, \eta_\mathcal{R} = 2.3, \eta_\mathcal{B} = 2.1, \gamma_\mathcal{R} = \gamma_\mathcal{B} = 2.4, \theta_{f,e}^\mathcal{B}(0) = \theta_{f,e}^\mathcal{R}(0) = 0.6 $. We consider two different values for $d_{max}$, $d_{max} = 100$ and $d_{max} = 1000$, for which the resulting equilibrium state (from the mean field model) is shown in Fig.\ref{fig:FigPropPA1}. For $d_{max} = 100$, consensus is achieved since in-group love ($\alpha = 0.15$) is considerably larger than out-group hate ($\beta = 0.001$) and the maximum number of enemies is $e_{max} = d_{max}-1 = 99$, which leads to all individuals leading to the most popular opinion, even if they have the highest possible number of enemies. However, when setting $d_{max}$ to a higher value ($d_{max} = 1000$), perfect consensus is not achieved. 

\subsubsection{Insights from Proposition \ref{prop:PA}}

In \textbf{Proposition} \ref{prop:PA}, we draw meaningful conclusions about the behavior of the system in a configuration model. First, we find sufficient conditions to reach perfect consensus. As seen in Fig. \ref{fig:FigPropPA1}, if we consider the mean field approximation model in a power-law degree distributed network with the initial conditions mentioned above, by increasing the maximum degree $d_{max}$ from $100$ to $1000$, we respectively transition from perfect consensus to partial consensus. In particular, in the case of $d_{max} = 1000$, individuals with one friend are leading towards the opposite choice (choice-1). This behavior is intuitive, as nodes with no friends under $d_{max} = 1000$  have a larger enemy influence than nodes with no friends under $d_{max} = 100$.

Since a key characteristic of many real-world social networks is power-law degree distribution \cite{HJeong_2003}, \cite{doi:10.1073/pnas.98.2.404}, this result sheds light on why perfect consensus remains unachievable in real-world. For example, in the past decade, the decision-making dynamics in American society have become divided between parties \cite{WebsterALanAPUSA}, which makes agreeing on decisions difficult, and therefore achieving perfect consensus unattainable \cite{KJamesAP},\cite{SLOTHUUS_BISGAARD_2021}.

\subsection{Numerical Results on the Dynamics on Configuration Model}

This subsection provides numerical simulations of the dynamics on the Configuration Model under the mean-field approximated trajectory to illustrate how the power-law degree distributions can lead to behaviors that are not emergent in the fully connected, stochastic block model, and Watts–Strogatz model, offering additional insights.

\begin{figure}[htbp]
    \centering
    
    \begin{subfigure}[b]{0.45\textwidth}
        \centering
        \includegraphics[width=\textwidth]{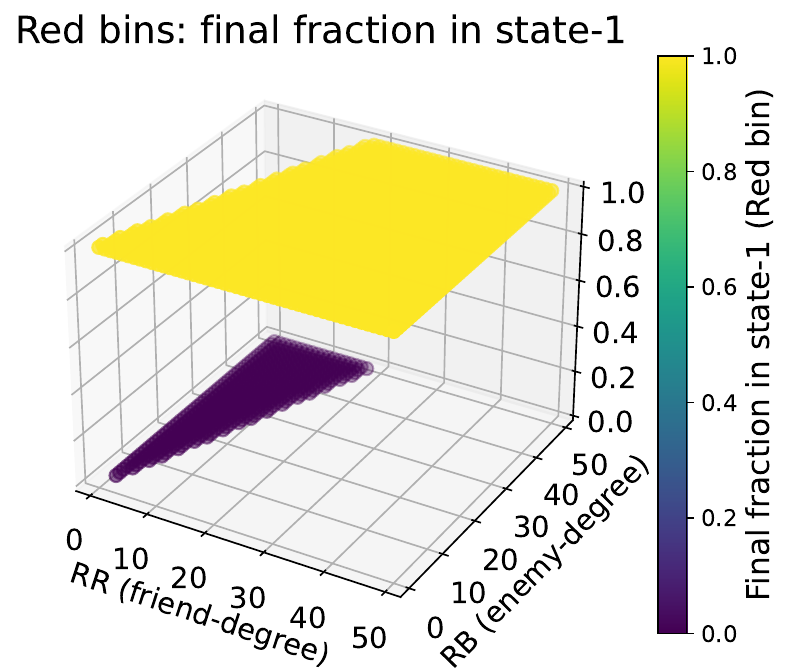}
        \caption{Fraction of red people on each bin that adopted choice-1 in the final state.}
        \label{fig:numredbin1}
    \end{subfigure}
    \hfill
    \begin{subfigure}[b]{0.45\textwidth}
        \centering
        \includegraphics[width=\textwidth]{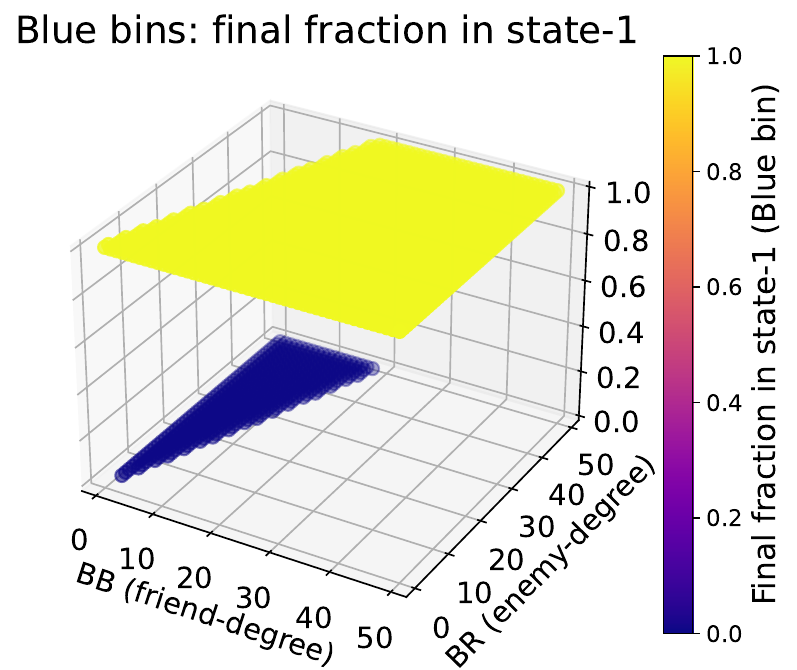}
        \caption{Fraction of blue people on each bin that adopted choice-1 in the final state.}
        \label{fig:numbluebin1}
    \end{subfigure}
    
    \vspace{1em} 
    \begin{subfigure}[b]{0.5\textwidth}
        \centering
        \includegraphics[width=\textwidth]{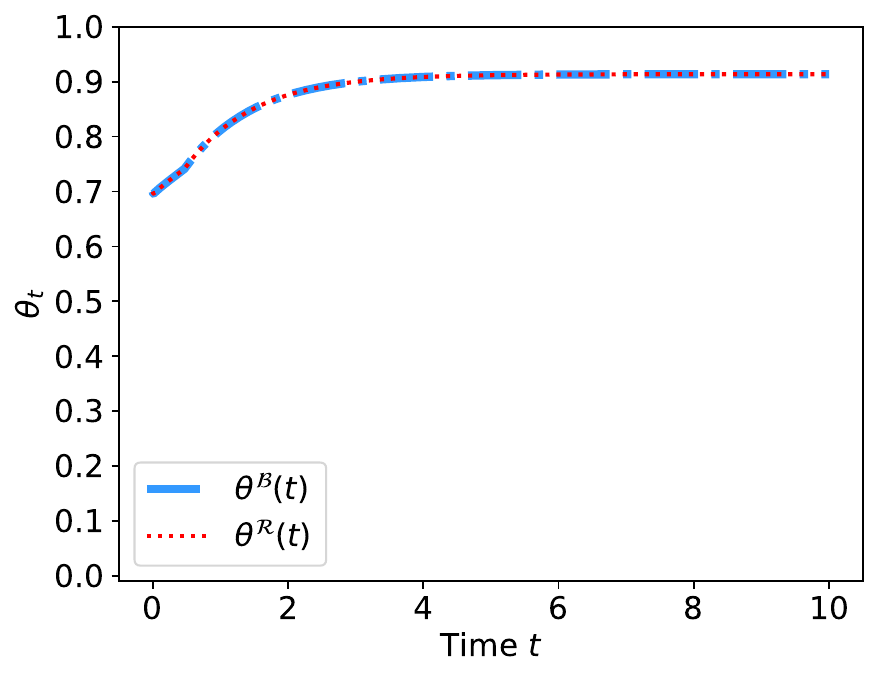}
        \caption{Overall trajectory of fraction of red and blue people that adopted choice-1.}
        \label{fig:sub3}
    \end{subfigure}
    
    \caption{Dynamics on Configuration Model for the following conditions: $d_{min} = 1, d_{max} = 50, \alpha =  \beta = 0.6, \eta_\mathcal{R} = \eta_\mathcal{B} = 2.1, \gamma_\mathcal{R} = \gamma_\mathcal{B} = 2.5, \theta_{f,e}^\mathcal{B}(0) = \theta_{f,e}^\mathcal{R}(0) = 0.7 $. }
    \label{fig:numcase1PA}
\end{figure}

Fig.~\ref{fig:numcase1PA} corresponds to the parameter configuration $d_{min} = 1, d_{max} = 50, \alpha =  \beta = 0.6, \eta_\mathcal{R} = \eta_\mathcal{B} = 2.1, \gamma_\mathcal{R} = \gamma_\mathcal{B} = 2.5, \theta_{f,e}^\mathcal{B}(0) = \theta_{f,e}^\mathcal{R}(0) = 0.7 $. Therefore, the two groups have the same parameter configurations and initial conditions. The mean-field dynamics show that each group gets split between the two choices. More specifically, individuals with more enemies than friends (i.e.,~the nodes in bins $f,e$ such that $e\gg f$) adopt choice-0 while others adopt choice-1 (which is the most popular choice initially in each bin). The intuition behind this observation is analogous to the intuition from Fig. \ref{fig:ApproxPAModel2}. This behavior leads to the following observation. Under an equal friend-enemy distribution ($\gamma_\mathcal{R} = \gamma_\mathcal{B}, \eta_\mathcal{R} = \eta_\mathcal{B}$), both groups exhibit a stronger and equal inclination towards the initially most popular decision choice-1 (see Fig.\ref{fig:numcase1PA}). This asymptotic behavior of the system differs from previous models studied, as there is no convergence to either choice $1$ or $0$. Under the given initial conditions, and in contrast to previously discussed models such as the fully connected case \cite{nettasinghe2025outgroup}, perfect consensus to the initially most popular decision (choice-1) would be anticipated.

\begin{figure}[htbp]
    \centering
    
    \begin{subfigure}[b]{0.45\textwidth}
        \centering
        \includegraphics[width=\textwidth]{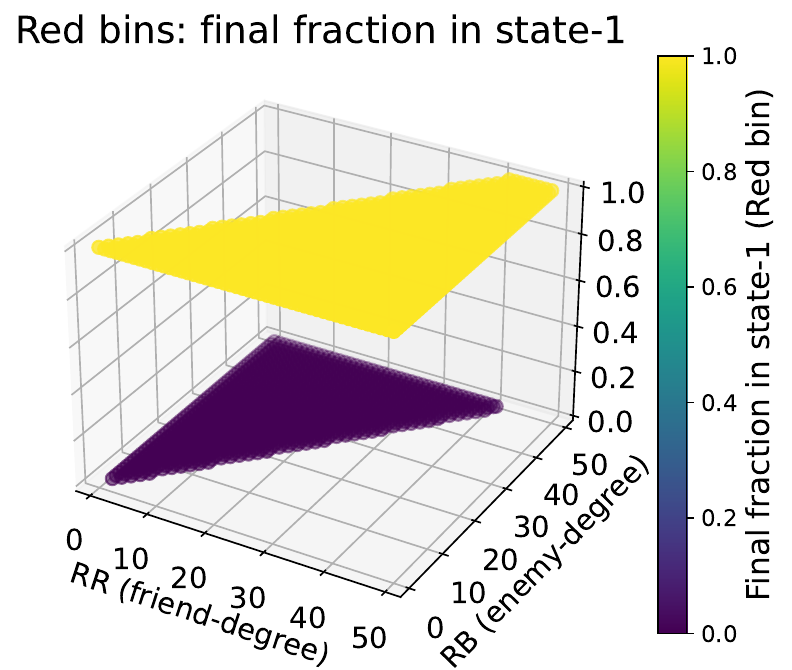}
        \caption{Fraction of red people on each bin that adopted choice-1 in the final state.}
        \label{fig:numredbin1}
    \end{subfigure}
    \hfill
    \begin{subfigure}[b]{0.45\textwidth}
        \centering
        \includegraphics[width=\textwidth]{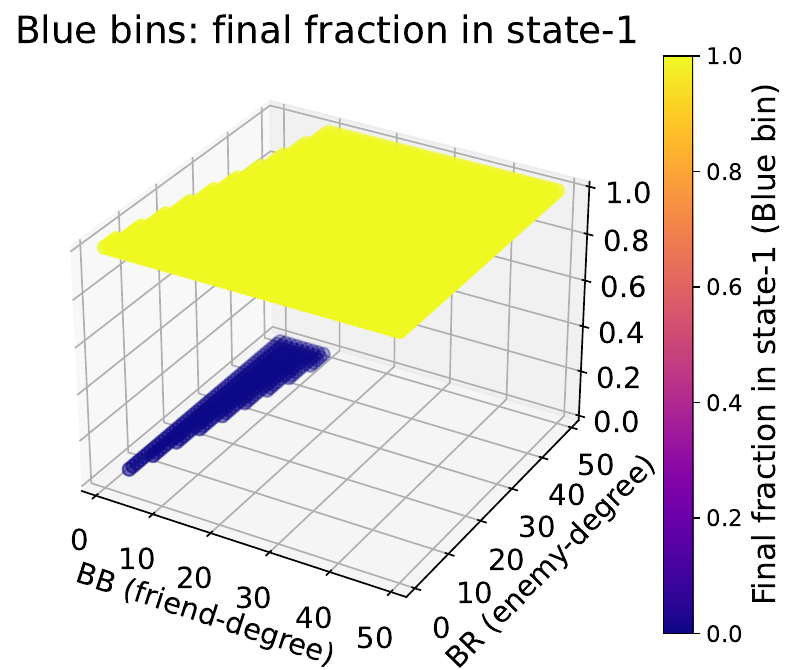}
        \caption{Fraction of blue people on each bin that adopted choice-1 in the final state.}
        \label{fig:numbluebin1}
    \end{subfigure}
    
    \vspace{1em} 
    \begin{subfigure}[b]{0.5\textwidth}
        \centering
        \includegraphics[width=\textwidth]{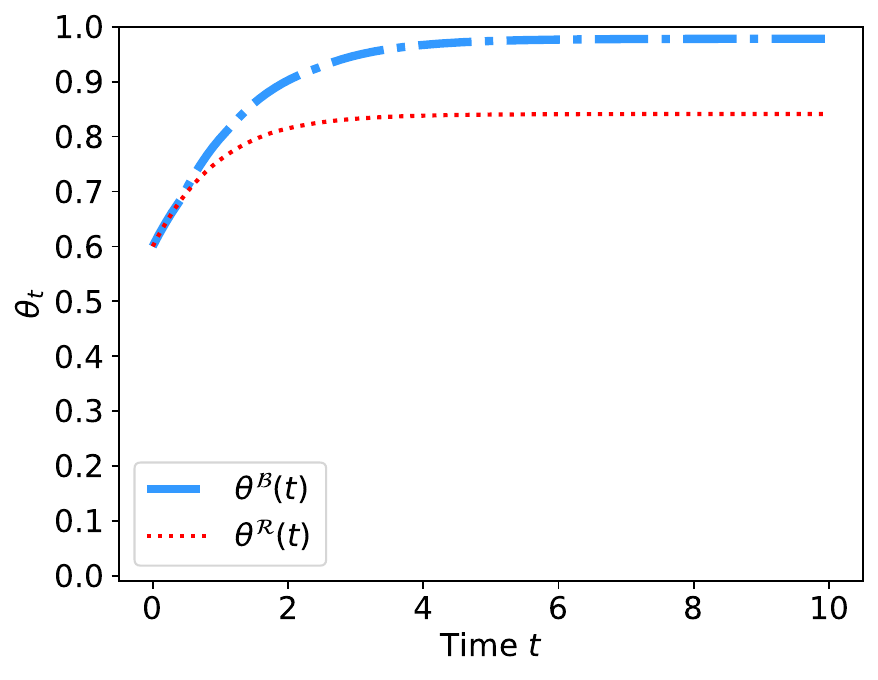}
        \caption{Overall trajectory of fraction of red and blue people that adopted choice-1.}
        \label{fig:sub3}
    \end{subfigure}
    
    \caption{Dynamics on Configuration Model for the following conditions: $d_{min} = 1, d_{max} = 50, \alpha =  \beta = 0.6, \eta_\mathcal{R} = 2.1, \eta_\mathcal{B} = 2.3, \gamma_\mathcal{R} = 2.7, \gamma_\mathcal{B} = 2.8, \theta_{f,e}^\mathcal{B}(0) = \theta_{f,e}^\mathcal{R}(0) = 0.6 $. }
    \label{fig:numcase2PA}
\end{figure}

Fig.~\ref{fig:numcase2PA} corresponds to the parameter configuration $d_{min} = 1, d_{max} = 50, \alpha =  \beta = 0.6, \eta_\mathcal{R} = 2.1, \eta_\mathcal{B} = 2.3, \gamma_\mathcal{R} = 2.7, \gamma_\mathcal{B} = 2.8, \theta_{f,e}^\mathcal{B}(0) = \theta_{f,e}^\mathcal{R}(0) = 0.6 $. Therefore, we have different exponent parameters with group independent initial conditions. The mean-field dynamics show that each group gets split between the two choices. We also note that both group present a different level of inclination for each choice. Similar to Fig. \ref{fig:numcase1PA}, the individuals with more enemies than friends (i.e.,~the nodes in bins $f,e$ such that $e\gg f$) adopt choice-0 while others adopt choice-1 (which is the most popular choice initially in each bin). This behavior leads to the following observation. Under a friend-enemy distributions such that $\gamma_\mathcal{R} < \gamma_\mathcal{B}, \eta_\mathcal{R} < \eta_\mathcal{B}$, both groups still exhibit a stronger inclination towards the initially most popular decision choice-1, with the inclination being stronger for the blue group than the red group (see Fig.\ref{fig:numcase2PA}). The intuition behind this behavior arises from the exponent parameters, as $\gamma_\mathcal{R} < \gamma_\mathcal{B}$ and $\eta_\mathcal{R} < \eta_B$, blue nodes connections are more evenly distributed than those from the red group. This asymptotic behavior of the system also differs from previous models studied, as each group exhibits a stronger inclination towards the initially most popular decision choice-1 in different levels.

\begin{figure}[htbp]
    \centering
    
    \begin{subfigure}[b]{0.45\textwidth}
        \centering
        \includegraphics[width=\textwidth]{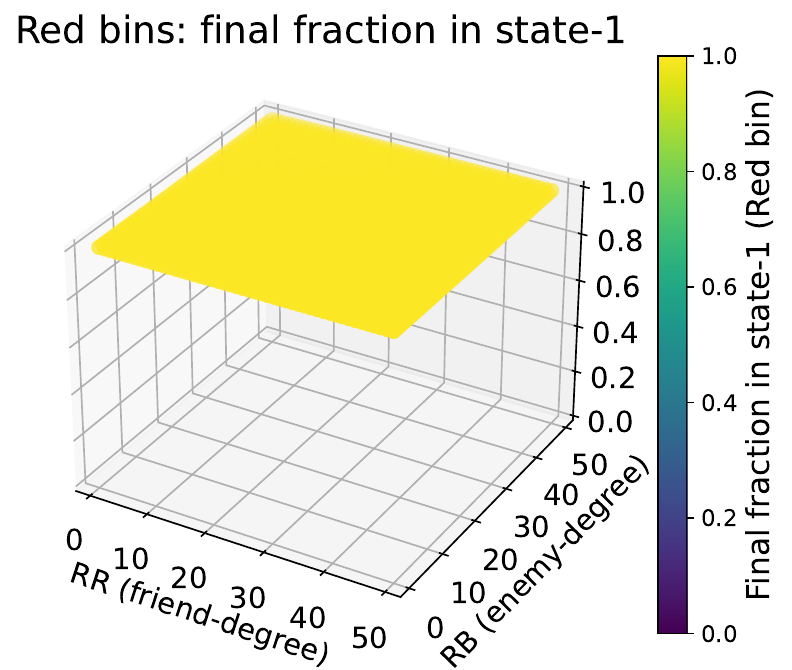}
        \caption{Fraction of red people on each bin that adopted choice-1 in the final state.}
        \label{fig:numredbin1}
    \end{subfigure}
    \hfill
    \begin{subfigure}[b]{0.45\textwidth}
        \centering
        \includegraphics[width=\textwidth]{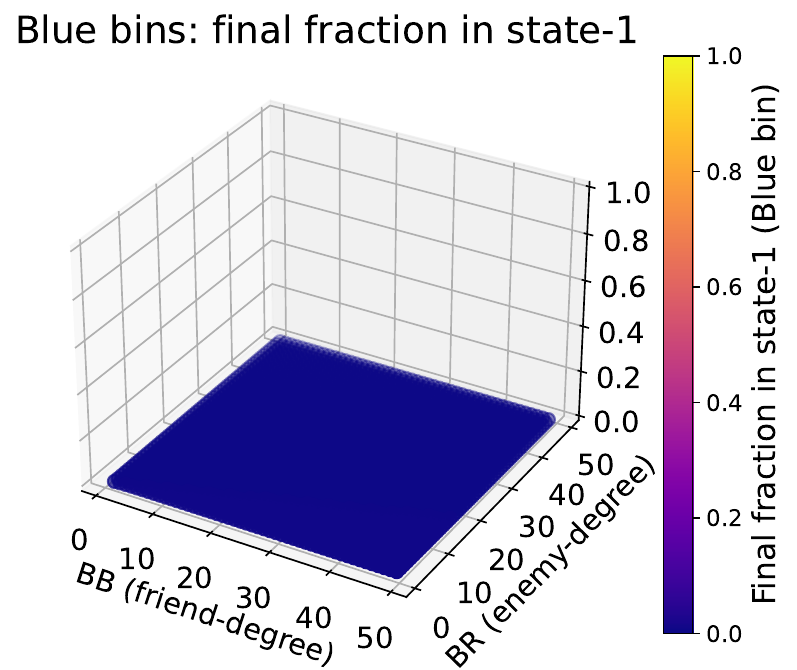}
        \caption{Fraction of blue people on each bin that adopted choice-1 in the final state.}
        \label{fig:numbluebin1}
    \end{subfigure}
    
    \vspace{1em} 
    \begin{subfigure}[b]{0.5\textwidth}
        \centering
        \includegraphics[width=\textwidth]{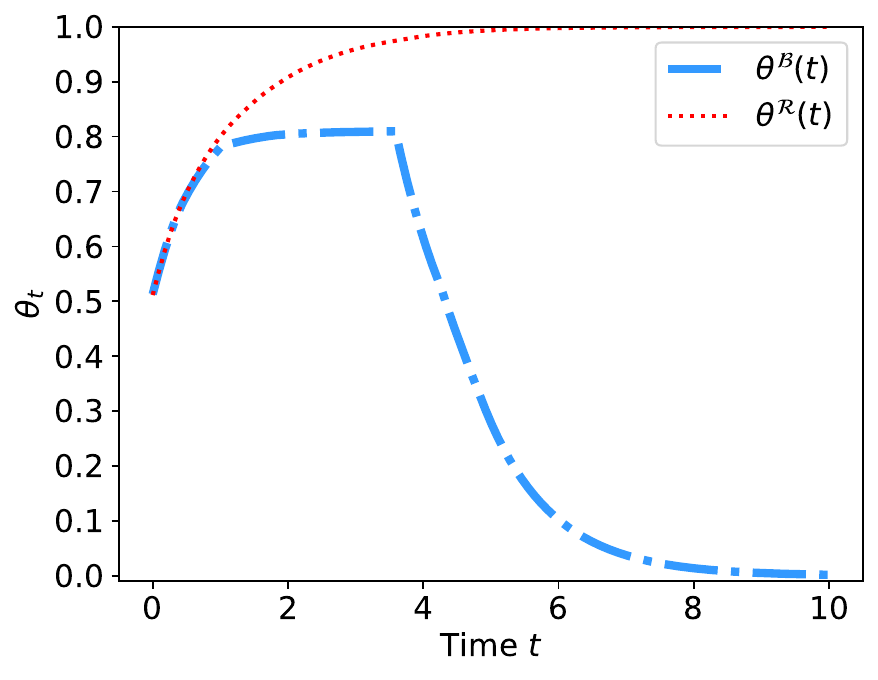}
        \caption{Overall trajectory of fraction of red and blue people that adopted choice-1.}
        \label{subfig:sub3}
    \end{subfigure}
    
    \caption{Dynamics on Configuration Model for the following conditions: $d_{min} = 1, d_{max} = 50, \alpha =  \beta = 0.6, \eta_\mathcal{R} = 2.5, \eta_\mathcal{B} = 2.4, \gamma_\mathcal{R} = 2.7, \gamma_\mathcal{B} = 2.6, \theta_{f,e}^\mathcal{B}(0) = \theta_{f,e}^\mathcal{R}(0) = \frac{f}{f+e} $. }
    \label{fig:numcase3PA}
\end{figure}

Fig.~\ref{fig:numcase3PA} corresponds to the parameter configuration $d_{min} = 1, d_{max} = 50, \alpha =  \beta = 0.6, \eta_\mathcal{R} = 2.5, \eta_\mathcal{B} = 2.4, \gamma_\mathcal{R} = 2.7, \gamma_\mathcal{B} = 2.6, \theta_{f,e}^\mathcal{B}(0) = \theta_{f,e}^\mathcal{R}(0) = \frac{f}{f+e} $. Therefore, we have different exponent parameters with bin dependent initial conditions. The mean-field dynamics show that in the long term, the red group adopts choice-1 while the blue group adopts choice-0. Under friend-enemy distributions such that $\gamma_\mathcal{R} > \gamma_\mathcal{B}, \eta_\mathcal{R}> \eta_\mathcal{B}$ and the initial state conditions are $\theta_{f,e}^\mathcal{B}(0) = \theta_{f,e}^\mathcal{R}(0) = \frac{f}{f+e} $, polarization arises. The intuition is the following. First, as $\theta_{f,e}^\mathcal{B}(0) = \theta_{f,e}^\mathcal{R}(0) = \frac{f}{f+e} $, nodes with more friends than enemies are initially more inclined toward choice-1 and nodes with more enemies than friends are initially more inclined towards choice-0. In addition, the red group has heavier-tailed friend degree distributions, which leads to the red group adopting choice-1 and the blue group adopting choice-0. A particularity of the dynamics in Fig. \ref{fig:numcase3PA} is the presence of a tipping point, visible in \ref{subfig:sub3}. Initially, both groups behaved identically until the differences in the friend–enemy distributions, i.e. the red group having heavier-tailed friend and enemy degree distributions, lead to the blue group converging to choice-0.

Overall, the examples discussed above show that affective polarization has richer dynamics in power-law networks (compared to previously studied fully connected, stochastic block model and watts-strogatz). This highlights the complexity of power-law distributed social networks compared to previously studied cases. 

To study the importance of the level of in-group love and out-group hate, we will simulate the system under the same conditions as in \ref{fig:numcase2PA} for different values of $\alpha$ and $\beta$.

\begin{figure}[htbp]
    \centering
    
    \begin{subfigure}[b]{0.32\textwidth}
        \centering
        \includegraphics[width=\textwidth]{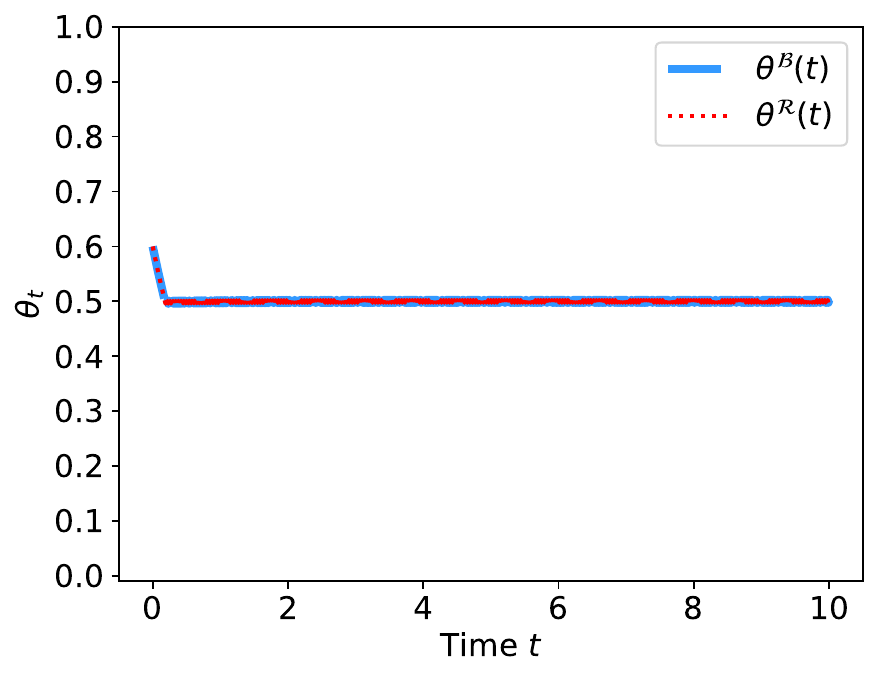}
        \caption{$\alpha << \beta$}
        \label{fig:hatew}
    \end{subfigure}
    \hfill
    \begin{subfigure}[b]{0.32\textwidth}
        \centering
        \includegraphics[width=\textwidth]{plotallN.pdf}
        \caption{$\alpha=\beta$}
        \label{fig:now}
    \end{subfigure}
    \hfill
    \begin{subfigure}[b]{0.32\textwidth}
        \centering
        \includegraphics[width=\textwidth]{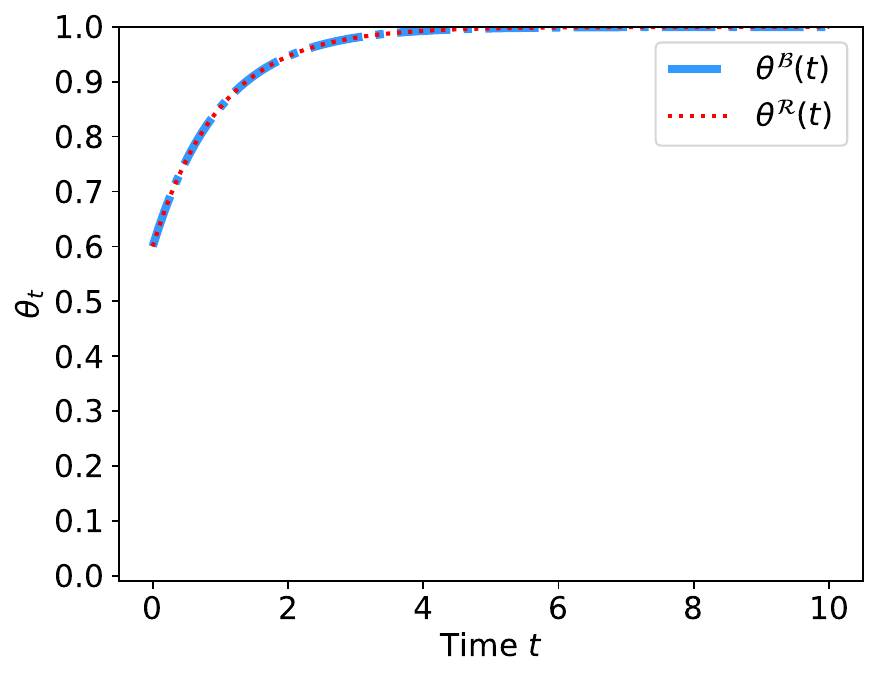}
        \caption{$\alpha >> \beta$}
        \label{fig:lovew}
    \end{subfigure}
    
    \caption{Overall trajectory of fraction of red and blue people that adopted choice-1 on Configuration Model for different values of $\alpha$ and $\beta$ under the following conditions: $d_{min} = 1, d_{max} = 50, \eta_\mathcal{R} = 2.1, \eta_\mathcal{B} = 2.3, \gamma_\mathcal{R} = 2.7, \gamma_\mathcal{B} = 2.8, \theta_{f,e}^\mathcal{B}(0) = \theta_{f,e}^\mathcal{R}(0) = 0.6$. The figure highlights the impact of the level of in-group love and out-group hate on the system. }
    \label{fig:numcase4PA}
\end{figure}

Fig. \ref{fig:numcase4PA} shows Fig. \ref{fig:numcase2PA} under different values of $\alpha$ and $\beta$. When out-group hate dominates over in-group love (Fig. \ref{fig:hatew}), the system reaches non-partisan polarization. However, when in-group love dominates over out-group hate (Fig. \ref{fig:lovew}), perfect consensus is reached. For both of these cases, the dominance leads to both groups having the same trajectories. The observed impact of the level of in-group love and out-group hate aligns with previous results from \cite{nettasinghe2025outgroup}.

\section{Experiments with Real World Datasets}
\label{sec:experiments}

In this section, we provide numerical experiments on the dynamics on the Configuration Model on a real-world social network dataset from Gowalla to explore how closely our theoretical model aligns with real-world data. The Gowalla dataset \cite{gowalla} contains $N = 196591$ nodes and $950327$ edges. To simulate the model, we randomly assigned groups to each node given that $Nr = \frac{N-1}{2}$. State initial conditions are also randomly assigned in each case. The exponents are approximated with the best power-law fit in which $\gamma_\mathcal{R} = 2.760,\gamma_\mathcal{B} = 2.645,\eta_\mathcal{R} = 2.747,\eta_\mathcal{B} = 2.874$. 

\begin{figure}[htbp]
    \centering
    \begin{subfigure}{0.48\textwidth}
        \centering
        \includegraphics[width=\linewidth]{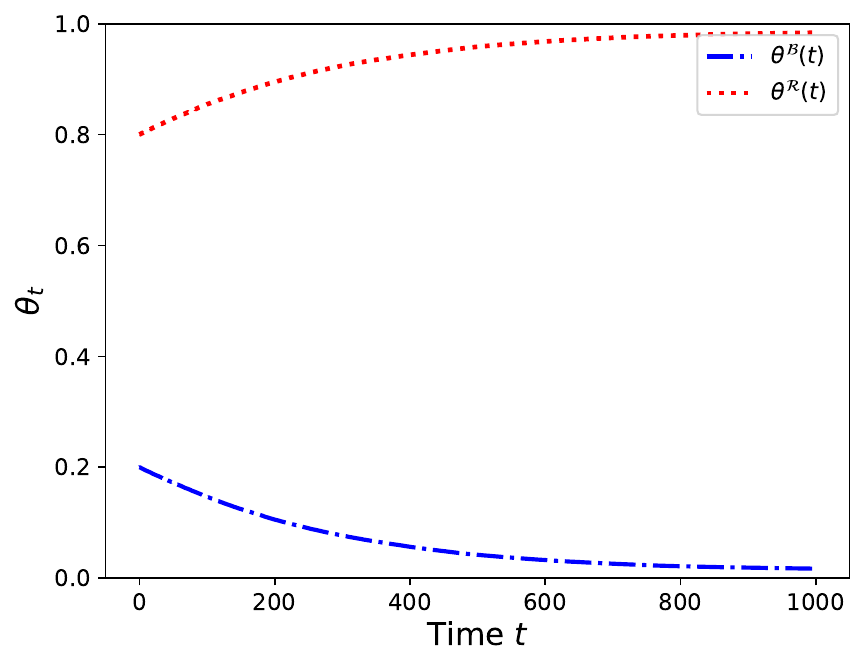}
        \caption{Discrete simulation on Gowalla.}
        \label{fig:gowallad1}
    \end{subfigure}
    \hfill
    \begin{subfigure}{0.48\textwidth}
        \centering
        \includegraphics[width=\linewidth]{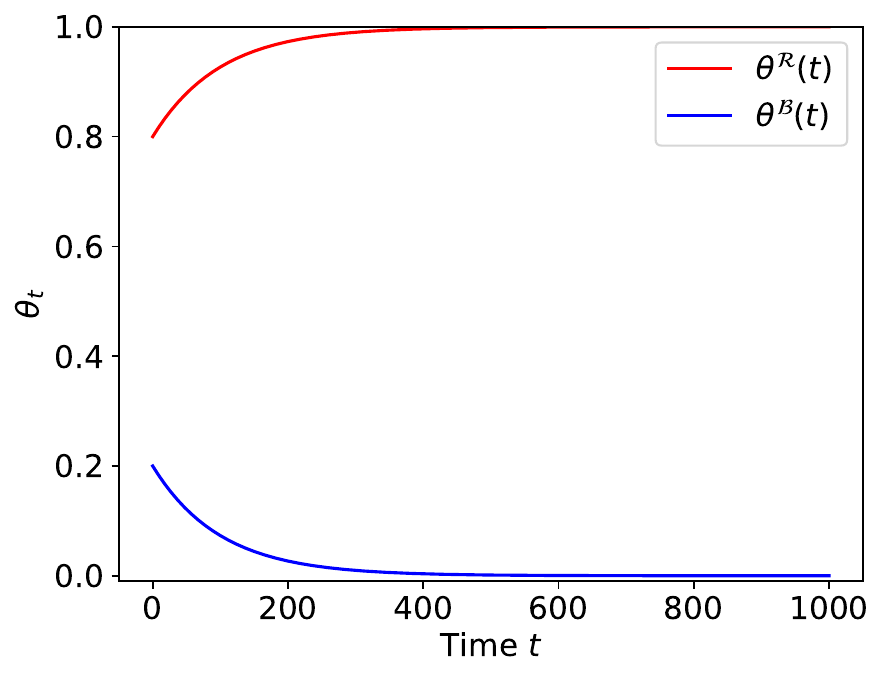}
        \caption{Theoretical trajectories.}
        \label{fig:gowallac1}
    \end{subfigure}

    \caption{Comparison between discrete simulation and theoretical trajectories for the Gowalla Dataset. For this case, $\alpha = \beta = 0.7, \theta_{f,e}^\mathcal{B}(0) = 0.2,  \theta_{f,e}^\mathcal{R}(0) = 0.8$.}
    \label{fig:gowalla1}
\end{figure}
Fig.~\ref{fig:gowalla1} corresponds to the dynamics of the Gowalla network under the parameters $\alpha = \beta = 0.7, \theta_{f,e}^\mathcal{B}(0) = 0.2 $ and $ \theta_{f,e}^\mathcal{R}(0) = 0.8$. Both dynamics show that the red group adopts choice-1 and the blue group adopts choice-0, leading to polarization. The observed dynamics are intuitive, as the level of in-group love and out-group hate are equal ($\alpha = \beta$) and each group adopts the most popular choice. The dynamics of the mean-field approximation (Fig. \ref{fig:gowallac1}) and the actual simulation (Fig. \ref{fig:gowallad1})  agree closely. In contrast to Fig.~\ref{fig:gowalla1}, we also explore Gowalla dynamics under group independent initial conditions and different values of in-group love and out-group hate. 
\begin{figure}[htbp]
    \centering
    \begin{subfigure}{0.48\textwidth}
        \centering
        \includegraphics[width=\linewidth]{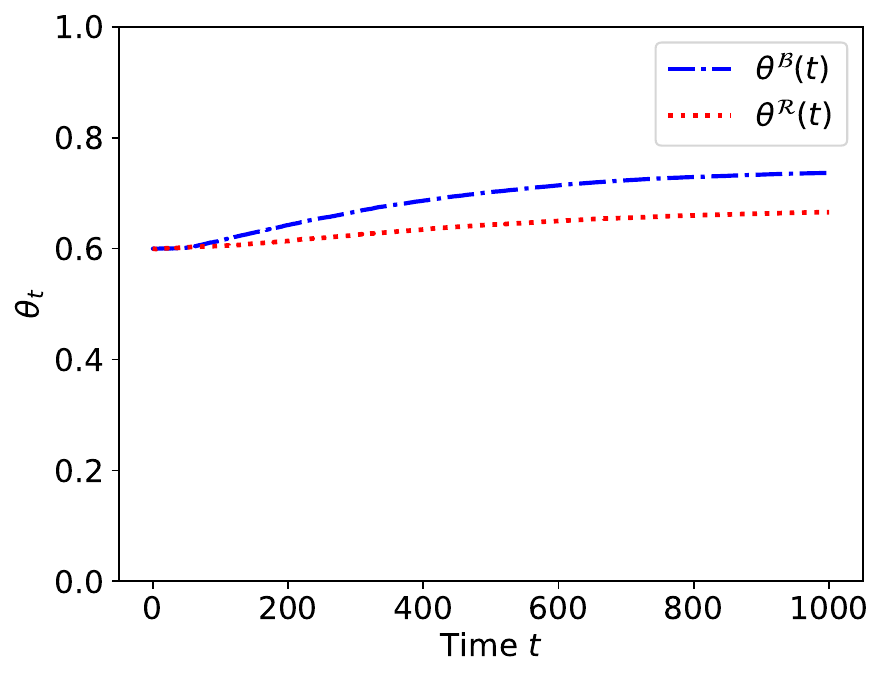}
        \caption{Discrete simulation on Gowalla.}
        \label{fig:gowallad2}
    \end{subfigure}
    \hfill
    \begin{subfigure}{0.48\textwidth}
        \centering
        \includegraphics[width=\linewidth]{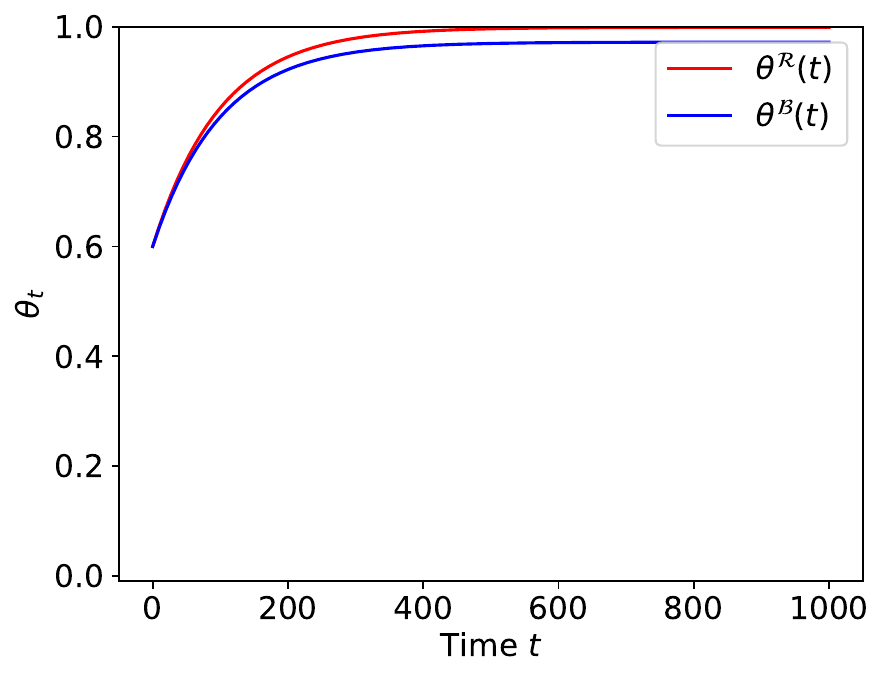}
        \caption{Theoretical trajectories.}
        \label{fig:gowallac2}
    \end{subfigure}

    \caption{Comparison between discrete simulation and theoretical trajectories for the Gowalla Dataset. For this case, $\alpha = 0.9, \beta = 0.1, \theta_{f,e}^\mathcal{B}(0) =  \theta_{f,e}^\mathcal{R}(0) = 0.6$.}
    \label{fig:gowalla2}
\end{figure}

Fig.~\ref{fig:gowalla2} corresponds to the dynamics of the Gowalla network under the parameters $\alpha = 0.9, \beta = 0.1 $ and $\theta_{f,e}^\mathcal{B}(0) =  \theta_{f,e}^\mathcal{R}(0) = 0.6$. Both dynamics show that, in the long term, for both groups, choice-1 remains the most popular opinion. As the level of in-group love is greater than the level of out-group hate ($\alpha >> \beta $), both groups are more inclined to the most popular choice (choice-1). However, the discrete simulation (Fig. \ref{fig:gowallad2}) shows that the blue group is more inclined to choice-1 than the red group, while the mean-field approximation (Fig. \ref{fig:gowallac2}) shows the opposite, i.e., the red group being more inclined to choice-1 than the blue group.
The agreement between the mean-field approximation and the actual simulation is visible but less so compared to Fig.~\ref{fig:gowalla1}.
This likely comes from a computational power limitation due to the finite number of time steps in the simulations and the finite size of the network. The discrete Gowalla simulation takes approximately 4 days to run. We can also consider the case of level of in-group love being less than the level of out-group hate.

\begin{figure}[htbp]
    \centering
    \begin{subfigure}{0.48\textwidth}
        \centering
        \includegraphics[width=\linewidth]{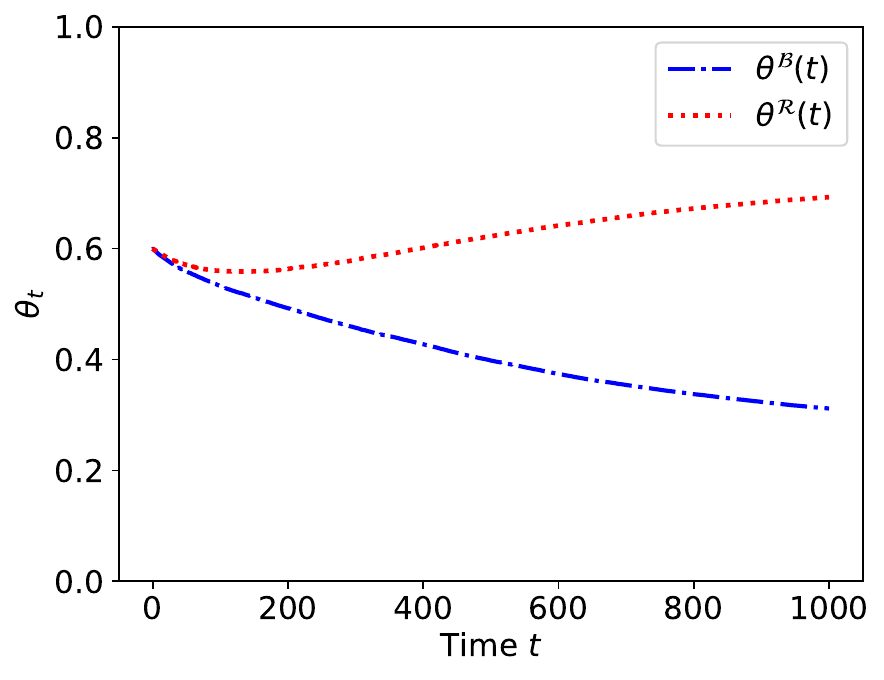}
        \caption{Discrete simulation on Gowalla.}
        \label{fig:gowallad3}
    \end{subfigure}
    \hfill
    \begin{subfigure}{0.48\textwidth}
        \centering
        \includegraphics[width=\linewidth]{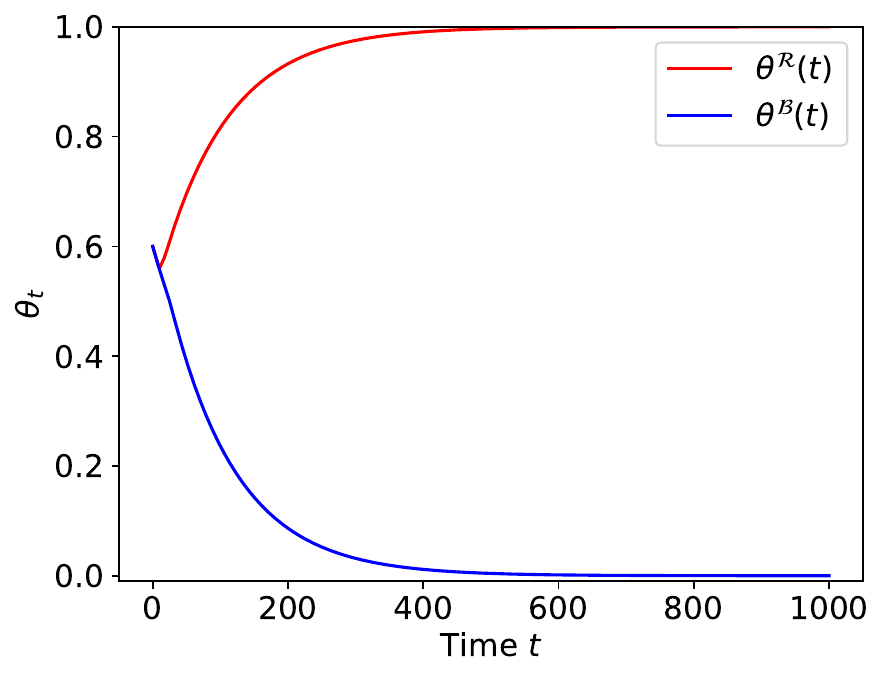}
        \caption{Theoretical trajectories.}
        \label{fig:gowallac3}
    \end{subfigure}

    \caption{Comparison between discrete simulation and theoretical trajectories for the Gowalla Dataset. For this case, $\alpha = 0.1, \beta = 0.9, \theta_{f,e}^\mathcal{B}(0) =  \theta_{f,e}^\mathcal{R}(0) = 0.6$.}
    \label{fig:gowalla3}
\end{figure}

Fig.~\ref{fig:gowalla3} corresponds to the dynamics of the Gowalla network under the parameters $\alpha = 0.1, \beta = 0.9 $ and $\theta_{f,e}^\mathcal{B}(0) =  \theta_{f,e}^\mathcal{R}(0) = 0.6$. The  mean-field approximation (Fig. \ref{fig:gowallac3}) shows that, in the long term, the red group adopts choice-1 while the blue group adopts choice-0. As the level of in-group love is less than the level of out-group hate ($\alpha << \beta $), polarization emerges. The discrete simulation (Fig. \ref{fig:gowallad3}) shows that over time, the red group increasingly inclines to choice-1 while the blue group decreasingly inclines to choice-0. The agreement between the mean-field approximation and the actual simulation is visible but less so compared to Fig.~\ref{fig:gowalla1}. This likely comes from a computational power limitation due to the finite number of time steps in the simulations.  

In summary, the comparisons of the discrete simulations and the mean-field approximation on the Gowalla network emphasize the usefulness of the configuration-model dynamics for real-world networks.

\section{Discussion}
\label{sec:discussion}

This article explored how the structure of the underlying social network affects the dynamics of opinions in an affectively polarized society. In particular, we extended the analysis of the model of opinion dynamics under affective polarization proposed in \cite{nettasinghe2025outgroup} to the small-world (Watts-Strogatz) networks and scale-free (power-law) networks. Our analysis provided mean-field approximations of the opinion dynamics under each model that relates the inherent properties of affective polarization (in-group love and out-group hate) to the properties of the network model. Our results reduce the need for computationally intensive simulations and facilitate an analytical understanding of the interplay between network structure and affective polarization. Numerical experiments show the alignment of mean-field dynamics and actual dynamics. 
 
For the Watts-Strogatz model, we find that a smaller rewiring constant facilitates consensus (where both parties eventually adopt the same decision) and increasing the rewiring constant leads to polarization. This finding suggests that \emph{smaller worlds} can amplify divisions, which agrees with prior research showing how digital media and breaking echo chambers facilitate polarization~\cite{nettasinghe2025group},\cite{polbackfire}.
Compared to the Watts-Strogatz model (and other previously studied models), the scale-free model shows a richer array of phenomena. Specifically, we show that heavy tailed degree distributions in scale-free networks make consensus difficult to achieve. This result deepens into why perfect consensus remains unattainable in the real-world \cite{deb54726-4c18-33be-aa6f-3f7e94124942}, \cite{consensusdryzek}. Moreover, from numerical simulations we observe that, for the dynamics of the scale-free network, convergence to partial consensus is possible (see Fig. \ref{fig:numcase1PA}). This behavior differs from previous models, where the only possible outcomes are perfect consensus, polarization, and non-partisan polarization.

The proposed models and their analysis have limitations that could extend this work in several directions for future research. In our findings, we assume fixed (static) levels of in-group love and out-group hate. Future work could look at how dynamic in-group love, out-group hate affect the dynamics. Previous models also rely on a binary society where individuals belong either to the red group or the blue group. In reality, group affiliation is often a spectrum, with some individuals strongly aligned, moderately aligned, or weakly aligned with a given group \cite{anes2024partyid},\cite{Elliott_2024}. Therefore, another meaningful direction is to consider a non-binary political party attribute.

\bibliographystyle{plain} 
\bibliography{references} 

\appendix
\section{Additional Details on Theoretical Results}
\label{sec:SupInfo}
\subsection{For $d\leq \sqrt{N} ln(N)$ on the ring lattice arrangement, the probability that a blue node has at least one red neighbor is zero for sufficiently large $N$. }
\label{subsec:SIdmin}
Let us remember that the way that each node is assign a political affiliation is in block order. The first $Nr$ nodes are red and the rest are blue. Moreover, in the ring lattice network, for an average degree $d$, every node is connected to its $\frac{d}{2}$ left and $\frac{d}{2}$ right neighbors. Therefore, in the array of the first $Nr$ nodes, only the first and last $\frac{d}{2}$ blue nodes can potentially have red neighbors. So, if $A$ is the event that a random blue node has at least one red neighbor then: 
$$ \mathcal{P}(A) \leq \frac{\frac{d}{2}+\frac{d}{2}}{N(1-r)} = \frac{d}{N(1-r)}$$ 
For $d \leq \sqrt{N} ln(N)$: 
$$\mathcal{P}(A) \leq \frac{d}{N(1-r)} \leq \frac{\sqrt{N} ln(N)}{N(1-r)} $$
$\text{As } N \to \infty, \frac{\sqrt{N} ln(N)}{N} \to 0$. \\
We can conclude that $\mathcal{P}(A)$ is zero for sufficiently large $N$ and $d \leq \sqrt{N} ln(N)$.

\subsection{Proof of Theorem \ref{TheoremSW}}
\label{subsec:proofth1}
\textbf{Stability of stationary states:}
Looking at the Jacobian matrix of the system in \ref{eq:systemSW}: for each
stationary state corresponding to cases 1-3, the Jacobian is a diagonal matrix with negative diagonal values, leading to stable steady states. For the non-partisan polarization (case 4), the fixed point is unstable. 

\textbf{Closed-form expressions:}
Let $\delta =0$ and $\theta^\mathcal{B}(0) = \theta^\mathcal{R}(0) = c$. 
Let's prove Case 1, i.e., suppose $ p < \frac{\alpha}{(\alpha+\beta) \cdot \max\{r,1-r\}}$.

If $c < 0.5$, then $$ \alpha(1 - pr)(2c - 1) - \beta pr(2c - 1) < 0$$ $$\alpha (1-p(1-r))(2c - 1) - \beta(1 - r)p(2c - 1) < 0$$
Hence at $t=0$, $p_{\theta}^\mathcal{R}(0 \to 1) = p_{\theta}^\mathcal{B}(0 \to 1) = 0$ and $p_{\theta}^\mathcal{R}(1 \to 0) = p_{\theta}^\mathcal{B}(1 \to 0) = 1$. Therefore, the dynamics are given by 
$$\dot{\theta}^\mathcal{B}(t) = -\theta^\mathcal{B}(t), \quad \dot{\theta}^\mathcal{R}(t) = -\theta^\mathcal{R}(t) $$
whose solution is $$\theta^\mathcal{B}(t) = \theta^\mathcal{R}(t) = ce^{-t} $$
which converges to $0$ asymptotically.

If $c > 0.5$, then $$ \alpha(1 - pr)(2c - 1) - \beta pr(2c - 1) > 0$$ $$\alpha (1-p(1-r))(2c - 1) - \beta(1 - r)p(2c - 1) > 0$$
Hence at $t=0$, $p_{\theta}^\mathcal{R}(0 \to 1) = p_{\theta}^\mathcal{B}(0 \to 1) = 1$ and $p_{\theta}^\mathcal{R}(1 \to 0) = p_{\theta}^\mathcal{B}(1 \to 0) = 0$. Therefore, the dynamics are given by 
$$\dot{\theta}^\mathcal{B}(t) =1 -\theta^\mathcal{B}(t), \quad \dot{\theta}^\mathcal{R}(t) = 1-\theta^\mathcal{R}(t) $$
whose solution is $$\theta^\mathcal{B}(t) = \theta^\mathcal{R}(t) = 1-(1-c)e^{-t} $$
which converges to $1$ asymptotically.

Proof of the other cases are analogous.
\subsection{Proof of Proposition \ref{prop:PA}}
\label{subsec:proofprop1}
 Let us consider the full system ${\theta}(t) = [\theta_{f,e}^\mathcal{B}(t), \theta_{f,e}^\mathcal{R}(t)]_{\{f,e=0,...,N-1 \}}$. Let $A = [1,1,1,...] $. We will prove under which conditions $A$ is a fixed point for the system in case 1. \\
$$A \text{ is a fixed point}  $$

$$ \Leftrightarrow \begin{bmatrix}
\dot{\theta}_{f,e}^\mathcal{B} \\
\dot{\theta}_{f,e}^\mathcal{R}
\end{bmatrix}
= 
\begin{bmatrix}
    0 \\
    0
\end{bmatrix}$$ given that $ {\theta}_{f,e}^\mathcal{R} = {\theta}_{f,e}^\mathcal{B} = 1$ for all $f,e = 0,...,N-1$. \\
$$ \Leftrightarrow \begin{bmatrix}
(1 - \theta_{f,e}^\mathcal{B})p_{f,e}^\mathcal{B}(0 \to 1) - \theta_{f,e}^\mathcal{B} p_{f,e}^\mathcal{B}(1 \to 0) = 0 \\
(1 - \theta_{f,e}^\mathcal{R})p_{f,e}^\mathcal{R}(0 \to 1) - \theta_{f,e}^\mathcal{R} p_{f,e}^\mathcal{R}(1 \to 0) = 0
\end{bmatrix}$$ for $ {\theta}_{f,e}^\mathcal{R} = {\theta}_{f,e}^\mathcal{B} = 1$ for all $f,e = 0,...,N-1$  \\ $$\Leftrightarrow p_{f,e}^\mathcal{B}(1 \to 0) = p_{f,e}^\mathcal{R}(1 \to 0) = 0 \text{ for all } e,f=0,...,N-1$$
$$ \Leftrightarrow \alpha f- \beta e > 0 \text{ for all } f,e = 0,...,N-1$$ 
$$ \Leftrightarrow \frac{\alpha}{\beta} > \frac{e}{f}\text{ for all } f,e = 0,...,N-1$$ 

$$ \Leftrightarrow \frac{\alpha}{\beta} > \max_{\substack{e,f=1,\dots,N-1}} \left\{ \frac{e}{f} \right\}$$ 

$$ \Leftrightarrow \frac{\alpha}{\beta} > \frac{e_{max}}{f_{min}}$$

  This inequality guarantees that the point $A = [1,1,1,...]$ is a fixed point of the system. Finding conditions for when $B = [0,0,0,...] $ is a fixed point is analogous and leads to the same inequality.
  
  $A$ and $B$ are stable steady states (point attractors). This can be formally verified by looking at the Jacobian: for case 1, the Jacobian is a diagonal matrix with negative diagonal values. If $c > 0.5$, then $\lim_{t \to \infty} \theta_{f,e}(t) = [1,1]^{'}$ for all $f,e$. If $c < 0.5$, then $\lim_{t \to \infty} \theta_{f,e}(t) = [0,0]^{'}$ for all $f,e$.

 The proof of case 2 consists of showing when $C = [\frac{1}{2},\frac{1}{2},\frac{1}{2},...] $ is a fixed point of the system, which is analogous to proving case 1. 

\section{Additional Details on Numerical Simulations}

\subsection{Construction of a graph with the configuration model}
\label{subsec:consCM}

To simulate the original discrete model, we need to create a graph with fixed initial conditions  $d_{max},d_{min}, \eta_\mathcal{R}, \eta_\mathcal{B}, \gamma_\mathcal{R}, \gamma_\mathcal{B}, \theta_{f,e}^\mathcal{B}(0)$ and $  \theta_{f,e}^\mathcal{R}(0)$. We first create $70000$ nodes for each group to get a total of $140000$ nodes. Then, to obtain the desired initial conditions $\theta_{f,e}^\mathcal{B}(0), \theta_{f,e}^\mathcal{R}(0)$ we randomly assigned groups and choices in the following way. A fraction $\theta_{f,e}^\mathcal{B}(0)$ of all blue nodes with $f$ friends and $e$ enemies are assigned choice-1 at time $t = 0$. The same process is followed for the red-group. However, for the other conditions, edges need to be carefully added. For this, we use the configuration model \cite{CMMolloy}. From  the probabilities of a red and blue node (respectively) of having $f$ friends and $e$ enemies, we can find the approximate degree sequence of connections between friends and enemies on the graph. For those degree sequences, we create stubs that are randomly paired to form edges. For each blue node, the number of friend half-edges (stubs) and enemy half-edges are sampled from the (unnormalized) probability distribution $f^{-\eta_\mathcal{B}}$ and the enemy degree is sampled from probability distribution $e^{-\gamma_\mathcal{B}}$. The friend and enemy half-edges for the red-group are assigned similarly. Then, the half-edges are paired randomly ensuring enemy edges connecting the two groups and friend edges staying within same group. This process may result in a random graph that contains self-loops and multi-edges. We discard any self-loops or multi-edges and do repairing until a valid edge is created.

The approximated $ \eta_\mathcal{R}, \eta_\mathcal{B}, \gamma_\mathcal{R}, \gamma_\mathcal{B}$ from the construction of the graph for the discrete trajectories in \ref{fig:ApproxPAModel2} and \ref{fig:ApproxPAModel2} are shown below.

\begin{figure}[h!]
    \centering

    \begin{minipage}[t]{0.48\textwidth}
        \centering
        \includegraphics[width=\linewidth]{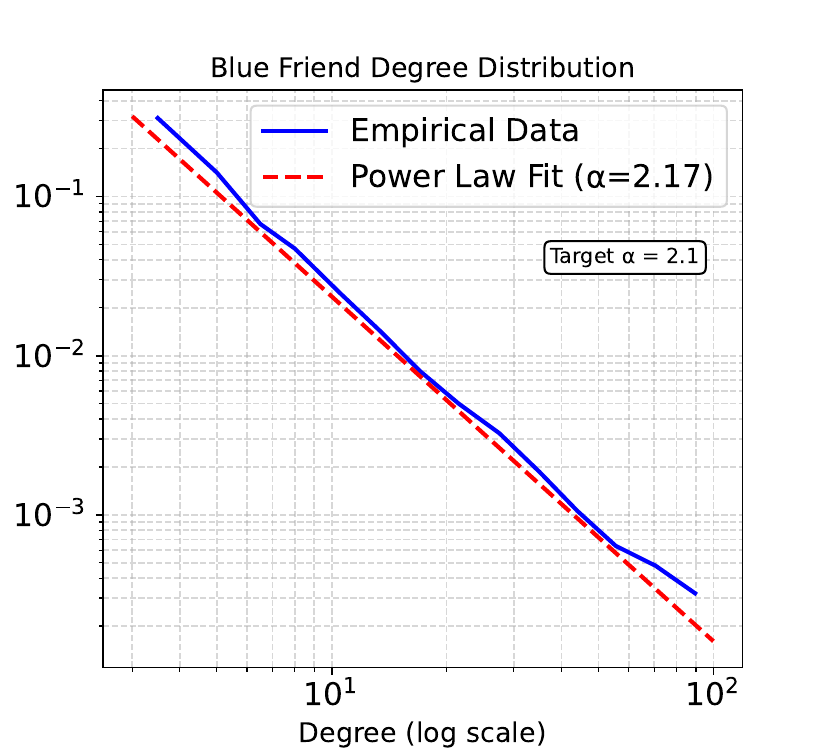}
        \caption*{(a)}
    \end{minipage}
    \hfill
    \begin{minipage}[t]{0.48\textwidth}
        \centering
        \includegraphics[width=\linewidth]{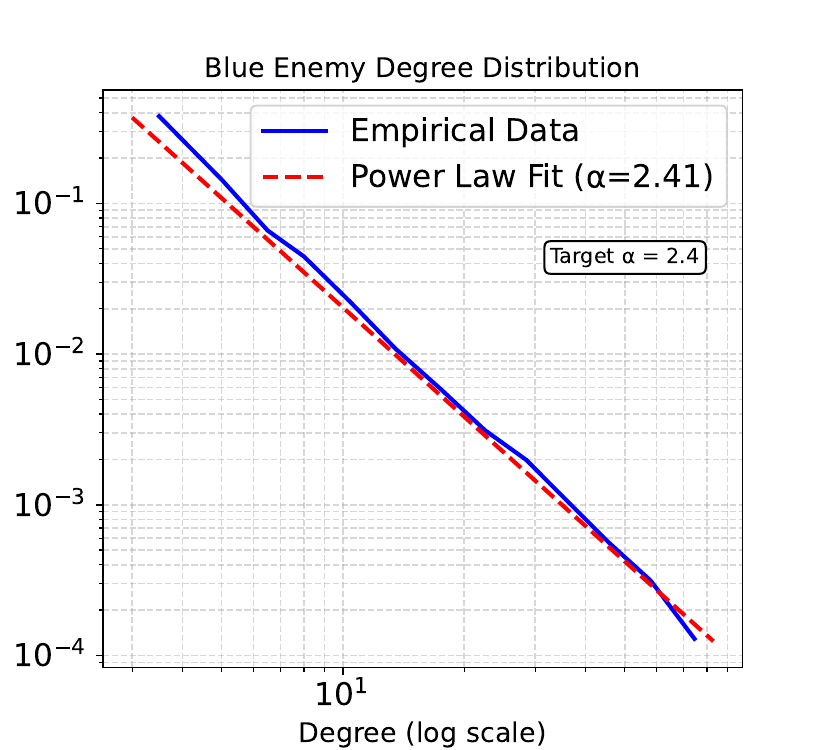}
        \caption*{(b)}
    \end{minipage}

    \vspace{0.3cm} 

    \begin{minipage}[t]{0.48\textwidth}
        \centering
        \includegraphics[width=\linewidth]{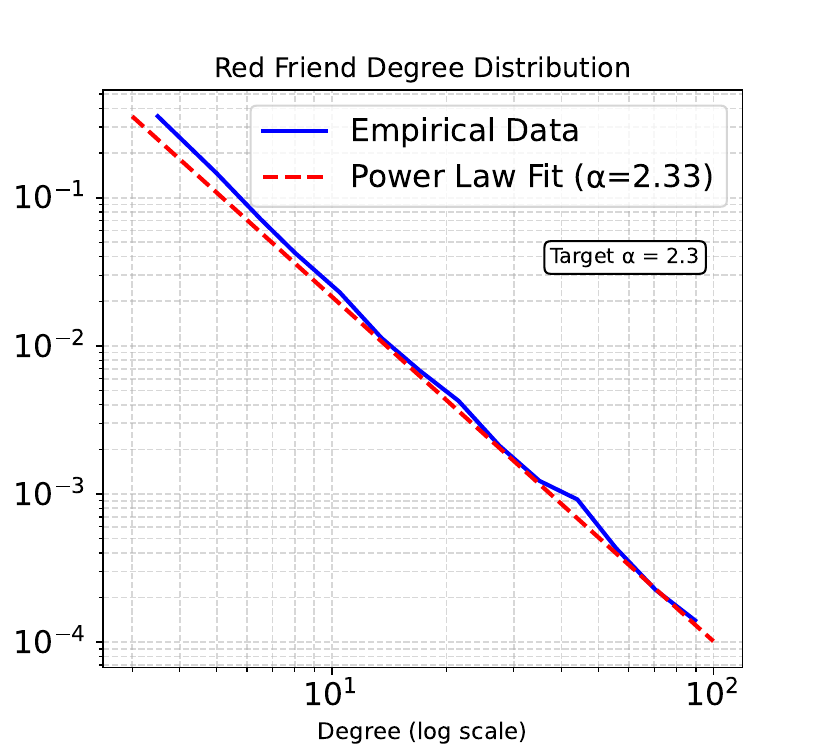}
        \caption*{(c)}
    \end{minipage}
    \hfill
    \begin{minipage}[t]{0.48\textwidth}
        \centering
        \includegraphics[width=\linewidth]{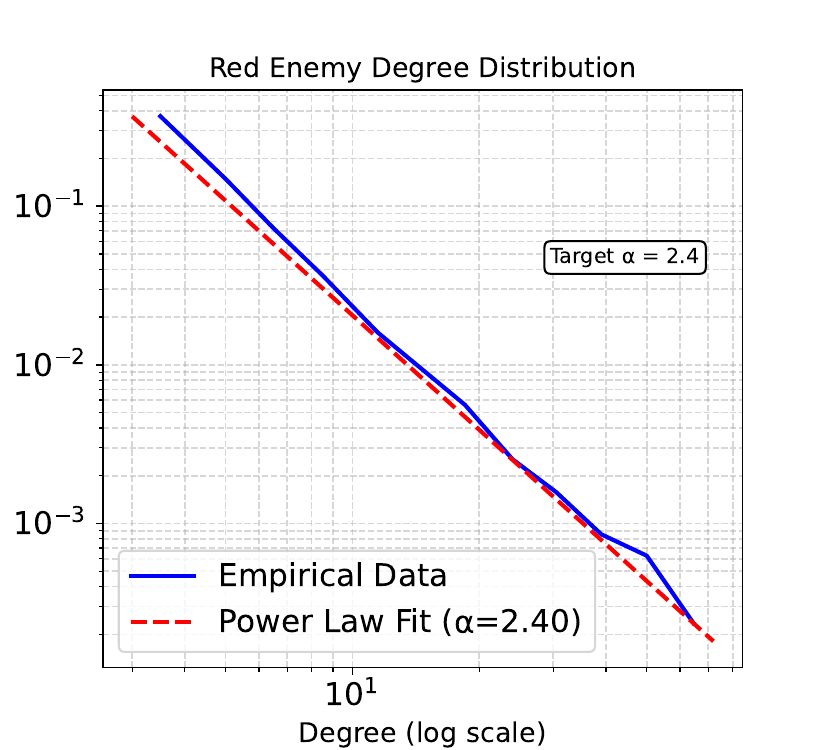}
        \caption*{(d)}
    \end{minipage}

    \caption{ Approximation of $ \eta_\mathcal{R}, \eta_\mathcal{B} , \gamma_\mathcal{R} , \gamma_\mathcal{B}$ on the construction of a graph to run the discrete model in Figure \ref{fig:ApproxPAModel1}. }
    \label{fig:EXPApproximation1}
\end{figure}

\begin{figure}[h!]
    \centering

    \begin{minipage}[t]{0.48\textwidth}
        \centering
        \includegraphics[width=\linewidth]{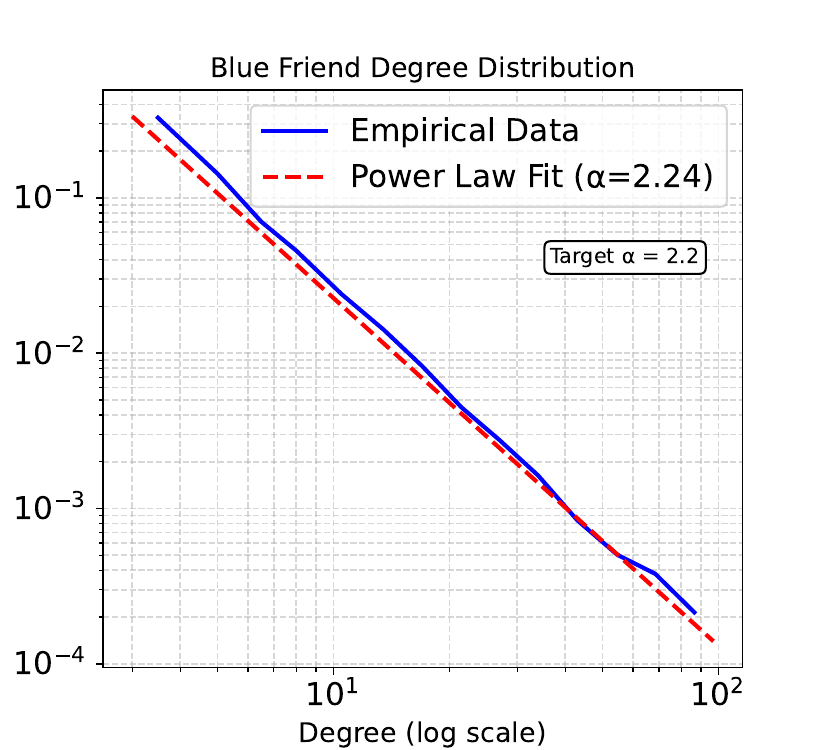}
        \caption*{(a)}
    \end{minipage}
    \hfill
    \begin{minipage}[t]{0.48\textwidth}
        \centering
        \includegraphics[width=\linewidth]{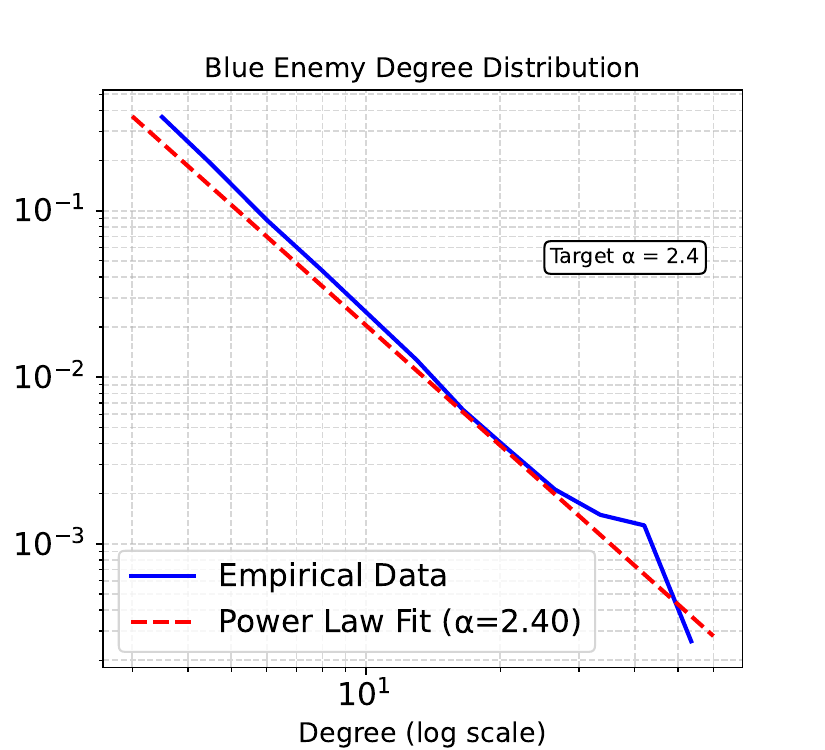}
        \caption*{(b)}
    \end{minipage}

    \vspace{0.3cm} 

    \begin{minipage}[t]{0.48\textwidth}
        \centering
        \includegraphics[width=\linewidth]{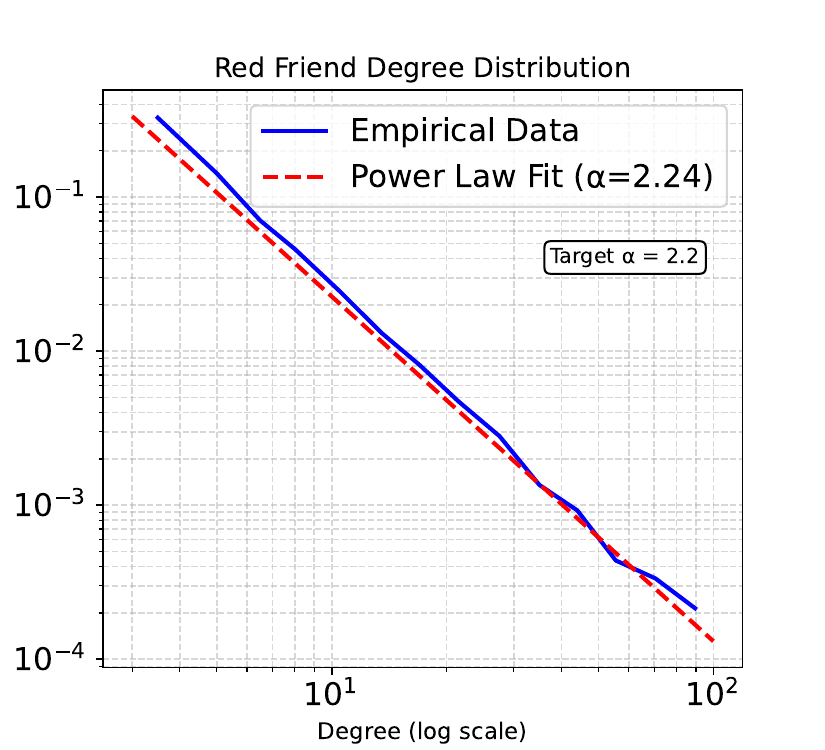}
        \caption*{(c)}
    \end{minipage}
    \hfill
    \begin{minipage}[t]{0.48\textwidth}
        \centering
        \includegraphics[width=\linewidth]{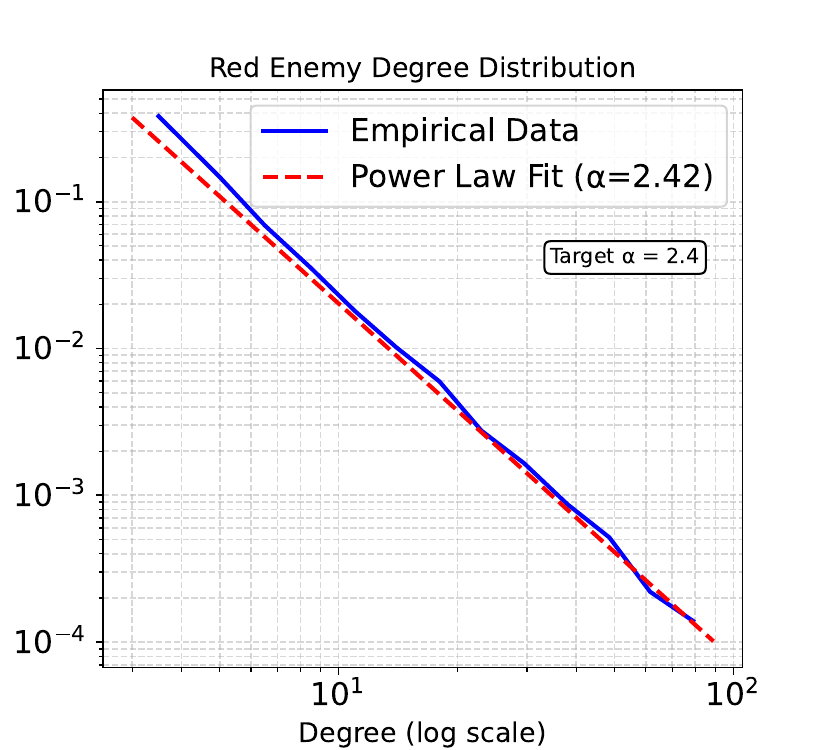}
        \caption*{(d)}
    \end{minipage}

    \caption{ Approximation of $ \eta_\mathcal{R}, \eta_\mathcal{B} , \gamma_\mathcal{R} , \gamma_\mathcal{B}$ on the construction of a graph to run the discrete model in Figure \ref{fig:ApproxPAModel2}. }
    \label{fig:EXPApproximation2}
\end{figure}

\end{document}